\PassOptionsToPackage{table,xcdraw}{xcolor}
\documentclass[11pt,a4paper]{article}
\pdfoutput=1
\usepackage{jheppub}
\usepackage{gensymb}
\DeclareSymbolFont{matha}{OML}{txmi}{m}{it}% txfonts
\DeclareMathSymbol{\varv}{\mathord}{matha}{118}
\usepackage{amsmath}
\usepackage{amssymb}
\usepackage{graphicx}
\usepackage{subfigure}
\usepackage{pstricks}
\usepackage{bm}
\usepackage{pbox}
\usepackage{placeins}
\usepackage{booktabs}
\usepackage[T1]{fontenc}
\usepackage{footnote}
\usepackage{pdfpages}
\usepackage{hhline}
\usepackage{multirow}
\usepackage{multicol}
\usepackage{enumitem}
\usepackage[toc,page]{appendix}
\allowdisplaybreaks
\usepackage{comment}
\usepackage{float}
\usepackage{slashed}
\usepackage{xcolor}
\usepackage{textcomp}
\usepackage{gensymb}
\usepackage{multirow}
\usepackage[numbers]{natbib}
\usepackage{notoccite}
\usepackage{cancel}
\usepackage{rotating}
\usepackage{tabularx}
\usepackage[labelfont=bf]{caption} % optional

\renewcommand{\thetable}{\Roman{table}}
\newcommand\scalemath[2]{\scalebox{#1}{\mbox{\ensuremath{\displaystyle #2}}}}

\definecolor{MyDarkBlue}{rgb}{0.1, 0.1, 0.8} 
\definecolor{MyLightBlue}{rgb}{0.22,0.51,0.9}
\definecolor{MyGreen}{rgb}{0.0, 0.5, 0.0}
\definecolor{BrickRed}{rgb}{0.8, 0.25, 0.33}

\RequirePackage{hyperref}
\hypersetup{colorlinks, citecolor=MyGreen,linkcolor=cyan, urlcolor=MyLightBlue}

\def\be{\begin{equation}}
\def\ee{\end{equation}}

\def\bea{\begin{eqnarray}}
\def\eea{\end{eqnarray}}

\definecolor{MyDarkBlue}{rgb}{0.1, 0.1, 0.8} 
\definecolor{MyLightBlue}{rgb}{0.22,0.51,0.9}
\definecolor{MyGreen}{rgb}{0.0, 0.5, 0.0}
\definecolor{BrickRed}{rgb}{0.8, 0.25, 0.33}

\RequirePackage{hyperref}
\hypersetup{colorlinks, citecolor=blue,linkcolor=cyan, urlcolor=MyLightBlue}

\makeatletter
\gdef\@fpheader{}
\makeatother
\begin{document}
\title{\bf 
One Phase to Rule Them All: Spontaneous CP Violation and Leptogenesis in SO(10)
}

\author[a]{K.S. Babu,}
\author[b]{Chee Sheng Fong,}
\author[c]{Shaikh Saad}  

\vspace{0.5cm}

\affiliation[a]{Department of Physics, Oklahoma State University, Stillwater, OK 74078, USA}

\affiliation[b]{Centro de Ciências Naturais e Humanas,
Universidade Federal do ABC, 09.210-170,\\ Santo André, SP, Brazil}

\affiliation[c]{Jožef Stefan Institute, Jamova 39, P.\ O.\ Box 3000, SI-1001 Ljubljana, Slovenia}

\emailAdd{babu@okstate.edu, sheng.fong@ufabc.edu.br, shaikh.saad@ijs.si}
%%%%%%%%%%%%%%%%%%%%%%%%%%%%%
\abstract{
We present a renormalizable $SO(10)$ grand unified theory with a \emph{minimal} Yukawa sector consisting of a $126_H$, a real $10_H$ and a real $120_H$, where $CP$ violation has a spontaneous origin. We show that the Yukawa sector in this setup, which consists of only 19 real parameters, is capable of simultaneously reproducing the observed fermion masses and mixings, including neutrino oscillations, as well as the baryon asymmetry of the Universe via thermal leptogenesis. In this framework, $CP$ is spontaneously broken when a $CP$-odd Higgs field $54_H$, used for GUT symmetry breaking, acquires a non-zero vacuum expectation value. All $CP$-violating phases, including the Dirac phases $\delta_\mathrm{CKM}$ and $\delta_\mathrm{PMNS}$, the neutrino Majorana phases, as well as those responsible for leptogenesis, arise solely from a single complex parameter in the Higgs potential. The proposed minimal setup predicts a \emph{normal} ordering of neutrino masses, with the atmospheric mixing angle $\theta_{23}$ preferred in the first octant and the leptonic $\delta_{\mathrm{PMNS}}$ lying in the range $(-38^\circ,\, +31^\circ)$. 
The fermion fits in our scenario further yield a strongly hierarchical mass spectrum for the three right-handed neutrinos, $(M_1,\,M_2,\,M_3) \sim \left(10^{5},\, 10^{12},\, 5\cdot 10^{14}\right)~\mathrm{GeV}$, which is shown to result in successful \emph{$N_2$-dominated} leptogenesis, consistent with current cosmological data.
}

%%%%%%%%%%%%%%%%%%%%%%%%%%%%%%%%%%%%%%%%%%
%%%%%%%%%%%%%%%%%%%%%%%%%%%%%%%%%%%%%%%%%%
\maketitle

%%%%%%%%%%%%%%%%%%%%%%%%%%%%%%%%%%%%%%%%%%%%%%%%%%
%%%%%%%%%%%%%%%%%%%%%%%%%%%%%%%%%%%%%%%%%%%%%%%%%%
\section{Introduction}
Understanding the origin of $CP$ violation, the observed pattern of fermion masses and mixings, including neutrino oscillation parameters, and the baryon asymmetry of the Universe remains a central challenge in particle physics. Grand Unified Theories (GUTs)~\cite{Pati:1974yy, Georgi:1974sy, Georgi:1974yf, Georgi:1974my, Fritzsch:1974nn} based on $SO(10)$ offer an exceptionally attractive framework for addressing these puzzles because all fermions of a given generation or family, including a right-handed neutrino, are unified in a single $16$-dimensional spinor representation, naturally incorporating the seesaw mechanism~\cite{Minkowski:1977sc,Yanagida:1979as,Glashow:1979nm,Gell-Mann:1979vob,Mohapatra:1979ia,Schechter:1980gr,Schechter:1981cv} for small neutrino masses.

While $SO(10)$ models have been extensively studied, fully realistic implementations often rely on  extended Yukawa structures, or additional symmetries, which can obscure predictivity. Of particular interest are models that achieve realistic fermion masses and mixings, including neutrino oscillations, along with thermal leptogenesis~\cite{Fukugita:1986hr} arising through the decays of heavy right-handed neutrinos and/or heavy Higgs triplet within a \emph{renormalizable} setup with a \emph{limited number} of parameters. Such constructions are highly constrained and can yield sharp and testable predictions. Fermion spectrum in such a setup has been analyzed in a variety of contexts. Refs.~\cite{Babu:1992ia,Joshipura:2011nn,Dueck:2013gca, Babu:2016cri,Babu:2020tnf} have studied both non-supersymmetric and supersymmetric $SO(10)$ models, whereas Refs.~\cite{Altarelli:2013aqa,Babu:2015bna,Babu:2016bmy,Ohlsson:2018qpt,Ohlsson:2019sja,Mummidi:2021anm,Saad:2022mzu,Haba:2023dvo, Kaladharan:2023zbr, Babu:2024ahk}   have focused on such spectrum in non-supersymmetric frameworks, while purely supersymmetric versions have been analyzed in Refs.~\cite{Babu:1995fp,Bajc:2001fe,Bajc:2002iw,Fukuyama:2002ch,Goh:2003sy,
Goh:2003hf,Dutta:2004hp,Bertolini:2004eq, Bertolini:2005qb,Babu:2005ia, Dutta:2005ni, Bertolini:2006pe, Aulakh:2006hs,Grimus:2006rk, Bajc:2008dc,Fukuyama:2015kra,Babu:2018tfi, Babu:2018qca, Mohapatra:2018biy, Mimura:2019yfi}.

An equally compelling puzzle in the flavor sector is an understanding of the origin of $CP$ violation. In the Standard Model (SM), $CP$ violation is an input parameter; in a unified framework, it is natural to ask whether it can emerge spontaneously from the vacuum structure. When implemented successfully, spontaneous $CP$ violation in $SO(10)$ can have the attractive feature that the same high-scale physics responsible for GUT symmetry breaking can simultaneously generate the complex phases that drive low-energy $CP$ violation in the quark and lepton sectors, as well as in leptogenesis. A major goal of this paper is to establish the feasibility of such a scheme.

In this work, we construct a renormalizable $SO(10)$ model with a \emph{minimal Yukawa sector} comprising a $126_H$, a real $10_H$, and a real $120_H$ representation. $CP$ symmetry is imposed at the Lagrangian level and is broken spontaneously when a $CP$-odd scalar $54_H$, breaks the $SO(10)$ symmetry down to an intermediate Pati--Salam symmetry.  Owing to $CP$ invariance, the Yukawa couplings are all real in this setup, implying that the Yukawa sector of the minimal model has only {\it 19 real parameters}, which we show can reproduce the 19 measured quantities in the quark and lepton masses and mixings, neutrino oscillation parameters and the baryon asymmetry of the Universe.  All $CP$-violating phases, including the Dirac phases in the quark and leptonic sectors, the two neutrino Majorana phases that appear in the effective Majorana mass for neutrinoless double beta decay 
%(and therefore the effective Majorana mass that relevant for neutrinoless double beta decay), 
and those relevant for leptogenesis, originate solely from a single complex entry in the Higgs potential. 
This, together with other real parameters in the Higgs potential, translate to \emph{four independent phases} in the vacuum expectation values (VEVs) in the Yukawa sector.
The resulting framework successfully reproduces all charged fermion masses, quark and lepton mixings, neutrino oscillation parameters, and the baryon asymmetry of the Universe via \emph{$N_2$-dominated} thermal leptogenesis~\cite{DiBari:2005st,Vives:2005ra,Blanchet:2006dq,Engelhard:2006yg}. In the context of  full $SO(10)$ frameworks, leptogenesis has been explored in  Refs.~\cite{Fong:2014gea,Mummidi:2021anm, Patel:2022xxu, Fu:2022lrn,Fu:2023mdu, Kaladharan:2023zbr, Babu:2024ahk,Fong:2025aya}, whereas  $SO(10)$-inspired scenarios are studied in Refs.~\cite{Buchmuller:1996pa, Nezri:2000pb,Buccella:2001tq,Branco:2002kt, Akhmedov:2003dg,DiBari:2008mp,DiBari:2010ux, Buccella:2012kc,DiBari:2013qja,DiBari:2014eya, DiBari:2015oca,DiBari:2017uka, Chianese:2018rnq,DiBari:2020plh,DiBari:2025zlv}. 

The high degree of constraint in our setup leads to several predictions:  a normal neutrino mass ordering; the atmospheric mixing angle $\theta_{23}$ likely in the first octant; $\delta_{\mathrm{PMNS}}$ favored in the range $(-38^\circ,\, +31^\circ)$; an extremely small lightest neutrino mass in the range $m_1 \sim (0.071, 0.127)~\mathrm{meV}$; and an effective neutrinoless double beta decay parameter in the range $m_{\beta\beta} \sim (1.20, 2.29)~\mathrm{meV}$.  These predictions can serve as tests of the model. 
Furthermore, the mass spectrum of the right-handed neutrinos $N_i\, (i=1,2,3)$ emerges as strongly hierarchical with $(M_1, M_2, M_3) \sim \left(10^{5},\, 10^{12},\, 5\times 10^{14}\right)~\mathrm{GeV}$, allowing a successful $N_2$-dominated leptogenesis.

This paper is organized as follows. In Sec.~\ref{sec:CP}, we present the model 
with an economical Yukawa sector, detailing the minimal Higgs content required to realize spontaneous $CP$ violation.
In Sec.~\ref{sec:yukawa}, we discuss the fermion mass matrices and the resulting low energy Yukawa couplings, perform a numerical fit to the observed fermion masses, mixings, and the baryon asymmetry parameter, and then employ a Markov chain Monte Carlo (MCMC) analysis to determine the likelihood predictions of several key observables. Finally, Sec.~\ref{sec:conclusion} summarizes our findings and outlines prospects for experimental tests. In appendices we present certain technical details. In Appendix~\ref{app:mass_matrix} we present details of deriving the Higgs doublet mass matrix while Appendix~\ref{app:bestfit} lists the benchmark fit parameters for fermion masses.

%%%%%%%%%%%%%%%%%%%%%%%%%%%%%%%%%%%%%%%%%%%%%%%%%%%%%%%%%%%%%
%%%%%%%%%%%%%%%%%%%%%%%%%%%%%%%%%%%%%%%%%%%%%%%%%%%%%%%%%%%%%
\section{Spontaneous CP Violation in SO(10)} \label{sec:CP}

%%%%%%%%%%%%%%%%%%%%%%%%%%%%%%%%%%%%%%%%%%%%%%%%%%%%%%%%%%%%%
\subsection{$CP$ and  properties of Yukawa couplings} 
Within the renormalizable setup, with no new fermions beyond the three families of chiral $16$s, the following fermion bilinears determine the Higgs representation needed to generate fermion masses in $SO(10)$:
\begin{equation}
16\times 16 = 10_s+ 120_a + 126_s.
\label{eq:sym}
\end{equation}
Here the subscripts $s$ and $a$ stand for symmetric and antisymmetric components. It is to be noted that the $10_H$ and the $120_H$ are real representations in $SO(10)$,  while the $126_H$ is complex.   For example, since $10_H$ is real, under the SM gauge group, it consists of the following conjugate fields:
\begin{align}
10_H = H(1,2,1/2) +  H^*(1,2,-1/2) + T(3,1,-1/3) +   T^*(\overline 3,1,1/3),
\end{align}
and similarly for the real $120_H$.  As demonstrated in Ref.~\cite{Babu:2016bmy}, the most economical Yukawa sector — defined as having the minimal number of Higgs fields and Yukawa parameters while remaining phenomenologically viable — consists of a single copy each of the real $10_H$ and the real $120_H$, along with one copy of the complex $126_H$ (we define $\overline{126}_H\equiv 126^*_H$). This leads to the following Yukawa  Lagrangian:  
\begin{equation}
\mathcal{L}_{yuk}= 16_F (y_{10} 10_H+y_{120} 120_H+y_{126} \overline{126}_H) 16_F + {\rm h.c.} \label{eq:yukawa}
\end{equation}
For brevity, the family indices are not explicitly shown.  The $3\times 3$ Yukawa coupling matrices $y_{10}, y_{126}$ are symmetric, whereas $y_{120}$ is antisymmetric, owing to the group property shown in Eq.~\eqref{eq:sym}.

Under $CP$ the fermions and the various scalar fields transform as~\cite{Grimus:1995zi,Grimus:2006rk}:\footnote{$10_H$, the fundamental representation, carries one group index. $120_H$ is a three index totally antisymmetric tensor, whereas $126_H$ is self-dual and completely antisymmetric five index tensor in the fundamental representation. We denote the SO(10) group indices as $p,q,...=1-10$.}  
\begin{align}
&16(x)\to  i C 16^*(\hat x),  
\\
& 10_p(x)\to \xi_p 10^*_p (\hat x),  \label{eq:10:CP}
\\
& 120_{pqr}(x)\to \xi_p \xi_q \xi_r 120_{pqr}^* (\hat x),   \label{eq:120:CP}
\\
& \overline{126}_{pqrst}(x)\to \xi_p \xi_q \xi_r \xi_s \xi_t \overline{126}_{pqr}^* (\hat x),   \label{eq:126:CP}
\end{align}
where $C$ is the charge conjugation matrix, $\hat x=(x^0,-\vec x)$, $\xi_p\equiv (-1)^{1+p}$.
%and $p, q, r, s, t$ are the $SO(10)$ indices. 
We will also need an
additional Higgs field for GUT symmetry breaking (discussed below) which we choose to be a $54_H$ (which is real and is denoted as a two index symmetric tensor). We assume the Higgs fields $10_H, 120_H, 126_H$ are $CP$-even, whereas $54_H$ is $CP$-odd.  The $CP$ transformation of the latter is therefore  given by
\begin{align}
& 54_{pq}(x)\to -\xi_p \xi_q 54_{pq}^* (\hat x).  \label{eq:54:CP}
\end{align}
Under the rules shown in  Eqs.~\eqref{eq:10:CP}-\eqref{eq:126:CP}, for a component field labeled as $S$, correspond to  (real$[S]$, imag$[S]$) $\to$(+real$[S]$, -imag$[S]$), whereas Eq.~\eqref{eq:54:CP} corresponds to  (real$[S]$, imag$[S]$) $\to$(-real$[S]$, +imag$[S]$). This is equivalent to a complex sccalar field transofroming under $CP$ to the corresponding complex conjugate field. The sign factors $\xi_p$ appearing in the transformation rules ensure this. Note also that $54_H$, contains a real SM singlet field $S$ which transforms  as $S\to -S$ under $CP$ as per Eq.~\eqref{eq:54:CP}, and thus will break $CP$  spontaneously when it acquires a VEV of order the GUT scale. This is the rationale behind the $CP$ assignment of the $54_H$ shown in Eq.~\eqref{eq:54:CP}. 
When the above-mentioned $CP$ transformation properties are combined with the symmetry of $y_{10}$ and $y_{126}$ and the antisymmetry of $y_{120}$, one ends up with \emph{real} Yukawa coupling matrices, namely,
\begin{align}
y_{10}^{ij}= (y_{10}^{ij})^*= y_{10}^{ji}, \quad
y_{120}^{ij}= (y_{120}^{ij})^*= -y_{120}^{ji}, \quad
y_{126}^{ij}= (y_{126}^{ij})^*= y_{126}^{ji},\label{eq:Yproperty}
\end{align}
where $i,j=1,2,3$ are the family indices.
The mass matrices arising from these real Yukawa matrices will nevertheless contain non-trivial phases, since the SM Higgs doublet will be a complex linear combination of the various Higgs doublets contained in $10_H$, $120_H$ and $126_H$. As we shall see, these complex parameters are dependent on a single complex term in the Higgs potential.

%%%%%%%%%%%%%%%%%%%%%%%%%%%%%%%%%%%%%%%%%%%%%%%%%%%%%%%%%%%%%
\subsection{$CP$ violation from GUT symmetry breaking} 
As will be shown in Sec.~\ref{sec:yukawa}, our model predicts a very high intermediate scale, 
$M_{\mathrm{int}} \sim 10^{14}~\mathrm{GeV}$. Such a large intermediate scale typically arises 
when $SO(10)$ is broken by a real $54_H$ Higgs representation. For a recent analysis of gauge coupling unification with this field, which yields a high-scale intermediate stage, see, for example, Ref.~\cite{Babu:2024ahk}. Therefore, for consistency, in this work we shall adopt the following symmetry-breaking chain:
\begin{align}
SO(10) 
&\xrightarrow[54_H]{M_\mathrm{GUT}} 
 SU(2)_{L} \times SU(2)_{R}\times SU(4)_{C}\times Z_2 \label{eq:SSB:01}
\\
&\xrightarrow[126_H]{M_\mathrm{int}} 
SU(3)_{c}\times SU(2)_{L} \times U(1)_{Y}   \label{eq:SSB:02}
\\ 
&\xrightarrow[10_H+120_H+126_H]{M_\mathrm{EW}} 
SU(3)_{c} \times U(1)_\mathrm{em}   \label{eq:SSB:03}
\end{align}
such that $M_\mathrm{GUT}> M_\mathrm{int}\gg M_\mathrm{EW}$. The $SO(10)$ gauge symmetry is first broken spontaneously to the Pati--Salam group supplemented by a discrete $Z_2$ symmetry, called C-parity, which interchanges the left and right components
of any representation, accompanied by charge conjugation~\cite{Kibble:1982dd}. 
Subsequently, the $126_H$ field acquires VEV of order $M_{\rm int}$, breaking the Pati--Salam symmetry down to the SM. 
The discrete symmetry  ensures the equality $g_{2L}^\mathrm{PS} = g_{2R}^\mathrm{PS}$, while the intermediate stage of symmetry breaking 
generates the masses of the heavy right-handed neutrinos.

To eliminate unwanted topological defects produced during subsequent phase transitions, we assume that cosmic inflation~\cite{Guth:1980zm,Albrecht:1982wi,Linde:1981mu,Linde:1983gd} occurs after the Pati--Salam symmetry is broken. As a result, all such defects, along with any baryon asymmetry generated by fields at or above the intermediate scale, are completely washed out.
As mentioned above, our choice of $54_H$ to break the GUT scale is dictated by the fermion mass fit. Consistency of the fermion mass fit requires an intermediate symmetry-breaking scale of order $10^{14}$ GeV~\cite{Babu:2016bmy}. If, instead of $54_H$, one were to choose, for example, a $210_H$ representation, the intermediate symmetry-breaking scale would be of order $10^{11}$ GeV~\cite{Babu:2016bmy}, assuming the validity of the extended survival hypothesis~\cite{Mohapatra:1982aq,Dimopoulos:1984ha}, in conflict with the requirements of the fermion sector.

To break $CP$ spontaneously, we assume that $54_H$ is $CP$-odd, namely,  $54_H \to - 54_H$, as explicitly written in Eq.~\eqref{eq:54:CP}.  As a consequence the cubic term $54_H^3$ is forbidden in the potential. Nevertheless, the GUT symmetry breaks spontaneously down to the Pati-Salam subgroup, with the following VEV structure: 
\begin{align}
    \langle 54_H \rangle = \omega_S \mathrm{diag}(1,1,1,1,-\frac{2}{3},-\frac{2}{3},-\frac{2}{3},-\frac{2}{3},-\frac{2}{3},-\frac{2}{3}),
    \label{eq:VEV54}
\end{align}
since, in addition to the quadratic term, the potential contains one trivial and one non-trivial quartic term in $54_H$. The intermediate Pati--Salam symmetry is further broken to the SM by the VEV of the \(126\), which we denote by \(\langle 126 \rangle = v_R\).

In the following, we demonstrate that, with the above set of Higgs fields along with their associated $CP$ properties, our model remarkably predicts the existence of a single complex entry in the Higgs doublet mass matrix originating at the high scale, which is responsible for generating all low-energy $CP$ violating phases.

%%%%%%%%%%%%%%%%%%%%%%%%%%%%%%%%%%%%%%%%%%%%%%%%%%%%%%%%%%%%%
\subsection{Origin of $CP$ violation at low energies} 

As mentioned above, since $CP$ is spontaneously broken in our setup, the Yukawa coupling matrices are all real. Thus, in order to reproduce the observed $CP$-violating quantities — such as the CKM phase in the quark sector and the baryon asymmetry parameter — the vacuum structure of the theory must be the source of $CP$ violation. Since the charged fermion masses arise from electroweak symmetry breaking, the origin of this high-scale $CP$ violation must be transmitted to the VEVs of the Higgs doublets.  The model contains five Higgs doublets, among which one linear combination is tuned to be light, identified as the SM Higgs doublet. The remaining four Higgs doublet fields have masses near the GUT scale. Since the SM Higgs doublet alone does not exhibit spontaneous $CP$ violation, $CP$ violation should arise only from the complex coefficients in the linear combination of the light Higgs doublet, which in the original basis (and not in the mass eigenbasis), can be interpreted as complex VEVs of the original Higgs doublet fields. We shall now construct the mass matrix of the Higgs doublets and show that only \emph{one} complex entry arising from spontaneous GUT symmetry breaking is sufficient to generate all the low energy phases.

First, note that the $10_H$ and $126_H$ representations each contains one up-type $(1,2,1/2)$ and one down-type $(1,2,-1/2)$ Higgs doublet (the quantum numbers refer to $SU(3)_c \times SU(2)_L \times U(1)_Y$ here), while the $120_H$ representation contains two such sets of submultiplets. 
Different from $126_H$, it is important to observe that the reality of $10_H$ and $120_H$ indicate that the corresponding up-type and down-type Higgs doublets are not independent but are related by complex conjugation.
Under the Pati--Salam subgroup $SU(2)_L \times SU(2)_R \times SU(4)_c$  of $SO(10)$ and the SM gauge symmetry these fields have the doublet decomposition given by
\begin{align}
10_H &\supset (2,2,1)  \supset  (1,2,1/2)_{(2,2,1)} + (1,2,-1/2)_{(2,2,1)}, \label{eq:10decompose:PS}
\\
120_H &\supset  (2,2,1) + (2,2,15) 
\nonumber \\&
\supset  (1,2,1/2)_{(2,2,1)} + (1,2,-1/2)_{(2,2,1)} + (1,2,1/2)_{(2,2,15)} + (1,2,-1/2)_{(2,2,15)},  \label{eq:120decompose:PS}
\\
126_H &\supset  (2,2,15) \supset    (1,2,1/2)_{(2,2,15)} + (1,2,-1/2)_{(2,2,15)}, \label{eq:126decompose:PS}
\end{align}
whereas under the  $SU(5) \times U(1)_X$ subgroup of $SO(10)$ and the SM symmetry they transform as
\begin{align}
10_H &\supset  (5,2)+(\overline 5,-2) \supset  (1,2,1/2)_{(5,2)} + (1,2,-1/2)_{(\overline 5,-2)}, \label{eq:10decompose:S5}
\\
120_H &\supset   (5,2)+(\overline 5,-2) + (45,2)+(\overline{45},-2) 
\nonumber\\&
\supset  (1,2,1/2)_{(5,2)} + (1,2,-1/2)_{(\overline 5,-2)} + (1,2,1/2)_{(45,2)} + (1,2,-1/2)_{(\overline{45},-2)},  \label{eq:120decompose:S5}
\\
126_H &\supset (\overline 5,-2)+  (45,2) +  \supset  (1,2,1/2)_{(45,2)} + (1,2,-1/2)_{(\overline 5,-2)}. \label{eq:126decompose:S5}
\end{align}
In the following, we adopt this compact notation $H^{R}_{M}$ to explicitly denote each Higgs doublet, where $R$ indicates the $SO(10)$ representation and $M$ labels the corresponding multiplet under a subgroup. The only exception to this notation occurs when referring to the Pati--Salam decomposition of $126_H$. Since both doublets arise from $(2,2,15)$, we will distinguish them by explicitly indicating whether they are of the up-type ($H_u$) or down-type Higgs doublet ($H_d$).   Since the $120_H$ contains two sets of up-type and two sets of down-type Higgs doublets, it is crucial to understand their decompositions under the two different subgroups of $SO(10)$. After performing an explicit computation (see Appendix~\ref{app:mass_matrix} for details), we obtain the following important transformation rule between these two bases:   
\begin{align}
&H^{120}_{(2,2,1)}= \frac{1}{2} H^{120}_{5} - \frac{\sqrt{3}}{2}  H^{120}_{45},  \label{eq:basis01}
\\
&H^{120}_{(2,2,15)}= \frac{\sqrt{3}}{2} H^{120}_{5} + \frac{1}{2}  H^{120}_{45}.   \label{eq:basis02}
\end{align}

We schematically summarize below the scalar potential terms relevant for computing the doublet mass matrix.
\begin{align}
&V \supset V_{10}+V_{120}+V_{126}+V_{10-126}+V_{120-126}+V_{10-120-126},
\end{align}
where 
\begin{align}
&V_{10} \supset 10^2_H+ (10_H 10_H)_{54} (54_H 54_H)_{54},  \label{eq:2.23}
\\
&V_{120} \supset 120^2_H+  (120_H 120_H)_{54} (54_H 54_H)_{54},   \label{eq:2.24}
\\
&V_{126} \supset 126_H \overline{126}_H+ \sum_{I=1}^4 ( 126_H^2 )_I ( \overline{126}_H^2 )_I
+ \underbrace{ \left( 126_H^2 54_H + h.c. \right) }_{\mathrm{purely\ imaginary}}
 + \underbrace{  \left( 126_H^2 54_H^2 +h.c. \right)  }_{\mathrm{purely\ real}} ,
\\
&V_{10-126} \supset (10_H 126_H)_{210} (126_H \overline{126}_H)_{210},
\\
&V_{120-126} \supset (120_H 120_H)_{210,770} (126_H \overline{126}_H)_{210,770}  
+  (126_H 126_H)_{1050} (120_H \overline{126}_H)_{\overline{1050}} ,
\\
&V_{10-120-126}  \supset    (10_H 120_H)_{45,210} (126_H \overline{126}_H)_{45,210}  .
\end{align}
In the above, $(10_H 10_H)_{54}$ denotes the appropriate contraction of the tensor product $10_H \otimes 10_H$ to yield the $54$ representation, with a similar convention applying to the other terms. The mass terms $10_H^2$ in Eq.~\eqref{eq:2.23} also contains contributions from $(10_H^2)_1(54_H^2)_1$ and $(10_H^2)_1(\overline{126}_H 126_H)_1$, and similarly for the $120_H^2$ mass term in Eq.~\eqref{eq:2.24}. 

At this stage, several important remarks are in order.
\begin{itemize}
\item Since our goal is to determine the complex nature of the Higgs doublet VEVs and establish whether they are independent, we only need the relative Clebsch–Gordan coefficients. Therefore, in the above potential, we have omitted the overall coefficients.

\item Intriguingly, the only source of complex parameters arises from the term 
\((126_H^2\,54_H + \text{h.c.})\), which is purely imaginary, and the term 
\((126_H^2\,54_H^2 + \text{h.c.})\), which is purely real. To see this, note
\begin{align}
&\mu 126_H 126_H 54_H +    \mu^* 126_H^* 126_H^* 54_H \xrightarrow{CP} -\mu 126_H^* 126_H^* 54_H - \mu^* 126_H 126_H 54_H,
\\
&\lambda 126_H 126_H 54^2_H +    \lambda^* 126_H^* 126_H^* 54^2_H \xrightarrow{CP} \lambda 126_H^* 126_H^* 54^2_H + \lambda^* 126_H 126_H 54^2_H.
\end{align}
These terms are \(CP\)-invariant if $\mu^*=-\mu$, i.e., this coefficient is purely imaginary,  
\(\mu = i\,\hat{\mu}\) (with \(\hat{\mu} \in \mathbb{R}\)), and  \(\lambda^*= \lambda\) is purely real.   We stress here that $\mu$ is the only fundamental complex parameter of the theory, with a phase of $\pi/2$.  Once the GUT symmetry is broken, in the Higgs doublet mass matrix, it will source a complex parameter of the form $\lambda \omega^2_s + i \hat \mu \omega_s$.

\item Due to \(CP\) invariance, all other terms must have real coefficients.

\item Although there are four invariants of the form $126_H^2 \overline{126}_H^2$, denoted as a sum $I = (1-4$), with $I$ taking values corresponding to the contractions yielding $(54, 1050, 2772, 4125)$, this only leads to two independent mass terms of the form $|H^{126}_u|^2$ and $|H^{126}_d|^2$.  This is because the singlet VEV of $126_H$ preserves $SU(5)$ which prevents mixing terms of the type $H_u^{126} H_d^{126}$, since that would be part of $(\overline{5}, -2) (45, 2)$ under $SU(5) \times U(1)$, which is not invariant under the surviving $SU(5)$.
These terms are then transformed into the Pati--Salam basis, which we adopt to write down the Higgs doublet mass matrix, using the relations given in 
Eqs.~\eqref{eq:basis01}--\eqref{eq:basis02}.

\item The term \((120_H\,120_H)_{54} \,(54_H\,54_H)_{54}\) in the above potential induces 
mass terms \(|H^{120}_{(2,2,1)}|^2\) and \(|H^{120}_{(2,2,15)}|^2\). 
By explicit computation, we find that the corresponding Clebsch–Gordan coefficients 
are related by a factor of \(9/17\) between these two terms, see Appendix~\ref{app:mass_matrix} for details of this derivation.

\item On the other hand, the terms 
\((120_H\,120_H)_{210,770} \,(126_H\,\overline{126}_H)_{210,770}\) 
induce mass terms of the form \(|H^{120}_{5}|^2\) and \(|H^{120}_{45}|^2\) with unrelated coefficients, which we transform into the Pati--Salam basis using
Eqs.~\eqref{eq:basis01}--\eqref{eq:basis02}.

\item Since $10_H$ contains only the $5 + \overline{5}$ decomposition under $SU(5)$ (and no $45 + \overline{45}$), see Eqs.~\eqref{eq:10decompose:PS}-\eqref{eq:10decompose:S5},  
the terms \((10_H\,120_H)_{45,210} \,(126_H\,\overline{126}_H)_{45,210}\)  
introduce mixing only between \(H^{10}_{5}\) and \(H^{120}_{\overline{5}}\),  
as well as between \(H^{10}_{\overline{5}}\) and \(H^{120}_{5}\).

\item Furthermore, the decompositions shown in Eqs.~\eqref{eq:120decompose:PS}-\eqref{eq:126decompose:PS} and Eqs.~\eqref{eq:120decompose:S5}-\eqref{eq:126decompose:S5} reveal that the term \((126_H 126_H)_{1050}\)   \((120_H \overline{126}_H)_{\overline{1050}}\) induces mixing terms of the form \(H^{120}_{\overline{5}} H^{\overline{126}}_{5}\) and \(H^{120}_{45} H^{\overline{126}}_{\overline{45}}\). We find that the relative Clebsch--Gordan coefficient of the latter term is \(1/\sqrt{3}\) compared to the former (see Appendix~\ref{app:mass_matrix}). 
\end{itemize}

Combining all the aforementioned results, we finally present the Higgs doublet mass matrix in the basis 
\begin{equation}
\vec{X} =
\left(H^{10}_{(2,2,1)}, H^{120}_{(2,2,1)}, H^{120}_{(2,2,15)}, H^{126}_{u}, \left(H^{126}_{d}\right)^*\right)^T
\end{equation}
written as $V \supset \vec{X}^\dagger \mathcal{M}^2_{(1,2,1/2)} \vec{X}$, with
\begin{align} \label{eq:doublet}
\mathcal{M}^2_{(1,2,1/2)}=
\scalemath{0.9}
{ 
\begin{pmatrix}
m_1^2 + \kappa_1 \omega_S^2 & \frac{\kappa_8 v_R^2}{2} & \frac{\sqrt{3}}{2} \kappa_8 v_R^2 & 0 & \kappa_9 v_R^2 \\
\frac{\kappa_8 v_R^2}{2} & m_{22}^2 & \frac{\sqrt{3}}{4} (\kappa_3 - \kappa_4) v_R^2 & -\frac{1}{2} \kappa_{10} v_R^2 & \frac{1}{2} \kappa_{10} v_R^2 \\
\frac{\sqrt{3}}{2} \kappa_8 v_R^2 & \frac{\sqrt{3}}{4} (\kappa_3 - \kappa_4) v_R^2 & m_{33}^2 & \frac{1}{2 \sqrt{3}} \kappa_{10} v_R^2 & \frac{\sqrt{3}}{2} \kappa_{10} v_R^2 \\
0 & -\frac{1}{2} \kappa_{10} v_R^2 & \frac{1}{2 \sqrt{3}} \kappa_{10} v_R^2 & m_3^2 + \kappa_6 v_R^2 & \color{red}(\lambda \omega_S^2 + i \hat\mu \omega_S) \\
\kappa_9 v_R^2 & \frac{1}{2} \kappa_{10} v_R^2 & \frac{\sqrt{3}}{2} \kappa_{10} v_R^2 & \color{red} (\lambda \omega_S^2 - i \hat\mu \omega_S) & m_3^2 + \kappa_7 v_R^2
\end{pmatrix}
},
\end{align}
where $m_{22}^2 \equiv m_2^2 + \kappa_2 \omega_S^2 + \frac{1}{4} \kappa_3 v_R^2 + \frac{3}{4} \kappa_4 v_R^2$ and $m_{33}^2 \equiv m_2^2 + \frac{17}{9} \kappa_2 \omega_S^2 + \frac{3}{4} \kappa_3 v_R^2 + \frac{1}{4} \kappa_4 v_R^2 $.
Here, appropriate coefficients are introduced, and the only source of $CP$ violation is highlighted in red. The mass parameters $m^2_{1,2,3}$, in addition to their bare masses, also receive contributions from products of trivial invariants. For example, for the $10_H$, this includes $(10_H^2)$, $(10_H^2)_1 (54_H^2)_1$, and $(10_H^2)_1 (126_H \overline{126}_H)_1$ terms. 

We assume that, apart from the SM Higgs doublet, which is a linear combination of the five doublets contained in $\vec{X}$, the remaining doublets have masses of order
$M_{H_{2,3,4,5}}^2 \sim M_H^2$ with $M_{\rm int} \lesssim M_H \lesssim M_{\rm GUT}$.
The light doublet is obtained by fine-tuning, which we accept. The determinant of the mass matrix in Eq.~\eqref{eq:doublet} must be nearly zero (more precisely of order $M_\mathrm{int}^8 M_W^2$ where $M_\mathrm{int}$ is the intermediate scale and $M_W$ represents the weak scale), with the zero mode composed of all five fields contained in $\vec{X}$. (Here we assume that the masses of these Higgs doublets are mildly tuned to $M_\mathrm{int}$. Tuning all masses to $M_\mathrm{int}$ may not be necessary when the two scales, $M_\mathrm{int}$ and $M_\mathrm{GUT}$, are very close to each other.)

Owing to the presence of the single complex entry in Eq.~\eqref{eq:doublet} the matrix that diagonalizes $\mathcal{M}^2_{(1,2,1/2)}$ must be unitary: 
\begin{equation}
U^\dagger \mathcal{M}^2_{(1,2,1/2)} U = {\rm diag.}\, (m_h^2, M_{H_2}^2, M_{H_3}^2, M_{H_4}^2, M_{H_5}^2)~.
\end{equation}
Here $m_h^2 \approx  -(88~\rm{GeV})^2$,  corresponding to the SM Higgs boson mass of 125 GeV, while all other squared masses are positive.  The unitary transformation acts on $\vec{X}$ as $\vec{X} = U \vec{X}'$, with $\vec{X}'$ denoting the mass eigenstates. This has the consequence that the VEVs of the Higgs doublets contained in $10_H, 120_H$ and $126_H$ are related to the electroweak VEV ($v = 174$ GeV) through the relations
\begin{eqnarray}
&~&\langle H^{10}_{(2,2,1)} \rangle \equiv v_{10} =  U_{11}\, v, ~~~\langle H^{120}_{(2,2,1)} \rangle \equiv v_{120}^{(1)} = U_{21} \, v, ~~~, 
\langle H^{120}_{(2,2,15)}\rangle  \equiv v_{120}^{(15)}= U_{31} \, v, \nonumber \\ 
&~&~~~~~~~~~~~~~~~~\langle H^{126}_{u}\rangle  \equiv v^u_{126} = U_{41} \, v,~~~ \langle (H^{126}_{d})^* \rangle \equiv (v_{126}^{(d)})^*  = U_{51}\,v    ~.
\label{eq:VEVs}
\end{eqnarray}
These relations, which apply to the up-type Higgs doublets, follow from the fact that among the five Higgs doublets only the SM Higgs acquires a nonzero VEV (equal to $v$).  

It is straightforward to derive the elements $U_{a1}\,(a=1-5)$ of the unitary matrix, which correspond to the eigenvector of the zero mode in Eq.~\eqref{eq:doublet} which can be written as
\begin{equation}
\sum_{b=1}^5 \left(\mathcal{M}^2_{(1,2,1/2)}\right)_{ab} U_{b1} = 0,~~~(a=1-5)~.
\end{equation}
These equations are analytically solvable for $U_{b1}$ appearing in Eq. (\ref{eq:VEVs}), but we shall not present their explicit form since they are lengthy. These equations involve all the parameters of the Higgs doublet mass matrix of Eq.~\eqref{eq:doublet}.  It can be easily verified that the coefficients $U_{b1}$ are complex with the phases induced by the single complex entry of Eq.~\eqref{eq:doublet}.  However, since the real parameters of Eq.~\eqref{eq:doublet} are relatively free, the phases of these electroweak VEVs can take arbitrary values, just as their magnitudes do. The magnitudes of the Higgs doublet VEVs should obey the condition
$v^2 = |v_{10}|^2 + |v_{120}^{(1)}|^2 + |v_{120}^{(15)}|^2 + |v_{126}^u|^2 + |v_{126}^d|^2 \simeq (174~{\rm GeV})^2$, which follows from the unitarity of $U$: $\sum_{b=1}^5 |U_{1b}|^2 = 1$.
In the numerical analysis of the next section we shall adopt a convention where $U_{41}$ is real, which can be achieved by a field redefinition of the light Higgs field.  This also shows that there are only \emph{four} physical phases in the Yukawa sector.

At this point, it is important to note that in the color triplet mass matrix, which are partners of the Higgs doublets, the \(\omega_S\) terms will appear with different Clebsch-Gordan coefficients compared to Eq.~\eqref{eq:doublet}. In addition, the color-triplet mass matrix is a \(6 \times 6\) matrix, compared to the \(5 \times 5\) matrix for the doublets, owing to an additional color-triplet contained in \(126_H\). Therefore, tuning the SM Higgs doublet to be light does not lead to any light color-triplets that would have resulted in rapid proton decay.

This concludes our proof that, for the fermion masses, each individual doublet VEV can be treated as an independent complex parameter, even though there is a single high-energy complex parameter behind the origin of all these phases.

Before leaving this section we wish to comment on some variations of the GUT symmetry breaking in our framework.  The fermion mass sector that is analyzed in the next section only depends on the presence of a minimal Yukawa sector consisting of a real $10_H$, a real $120_H$ and a complex $126_H$, along with spontaneous $CP$ violation arising through the complex VEVs for the Higgs doublets.  The GUT symmetry, as well as $CP$ symmetry, is allowed to be broken by Higgs fields other than the $54_H$. In all cases that we have analyzed the results on fermion fits and leptogenesis  will hold, as elaborated below.

One could envision that the $54_H$ that breaks the GUT symmetry is $CP$-even, in which case an $SO(10)$ singlet Higgs $S$, which is $CP$-odd may break $CP$ spontaneously.  The coefficients of the couplings $(126_H)^2 54_H$ and $(126_H)^2 (54_H)^2$ are both real in this case, while the coupling $(126_H)^2 54_H S$ will be purely imaginary.   This will lead to the same form of the Higgs doublet mass matrix as in Eq.~\eqref{eq:doublet}.  However, this scenario involves an SO(10) group singlet and therefore may not be very appealing.  If a $210_H$ is used in place of the $54_H$, as noted before, the intermediate scale will be of order $10^{11}$ GeV, assuming the validity of extended survival hypothesis.  One could consider deviating from this hypothesis significantly, for example, by lowering the mass of a scalar multiplet $(3,1,10)$ under the Pati-Salam symmetry to $M_{int}$, although it is not needed for symmetry breaking, in which case $M_{int}$ can be made close to $10^{14}$ GeV.  In this case, there are two purely imaginary couplings in the Higgs potential, viz., $10_H 126_H 210_H$ and $120_H 126_H 210_H$, with all other couplings being real.  Even in this setup our analysis of fermion masses and mixing as well as leptogenesis will go through. The Higgs doublet mass matrix in this case will be $6 \times 6$ dimensional, owing to the presence of a doublet in $210_H$, but the VEVs of the doublets can have arbitrary magnitudes and phases in this case as well.

%%%%%%%%%%%%%%%%%%%%%%%%%%%%%%%%%%%%%%%%%%%%%%%%%%
%%%%%%%%%%%%%%%%%%%%%%%%%%%%%%%%%%%%%%%%%%%%%%%%%%
\section{Minimal Yukawa Sector}\label{sec:yukawa} 
%%%%%%%%%%%%%%%%%%%%%%%%%%%%%%%%%%%%%%%%%%%%%%%%%%%%%%%%%%%%%
\subsection{Fermion mass matrices} 
The fermion mass matrices derived from the Yukawa Lagrangian given in Eq.~\eqref{eq:yukawa} are as follows~\cite{Babu:2016bmy}:
\begin{align}
&M_U= v_{10}y_{10}+v^u_{126}y_{126}+(v^{(1)}_{120}+v^{(15)}_{120})y_{120},\label{Mu}\\
&M_D= v^{\ast}_{10}y_{10}+v^d_{126}y_{126}+(v^{(1)\ast}_{120}+v^{(15)\ast}_{120})y_{120},\\
&M_E= v^{\ast}_{10}y_{10}-3v^d_{126}y_{126}+(v^{(1)\ast}_{120}-3v^{(15)\ast}_{120})y_{120},\\
&M_{\nu_D}= v_{10}y_{10}-3v^u_{126}y_{126}+(v^{(1)}_{120}-3v^{(15)}_{120})y_{120},\\
&M_{\nu_{R}}=v_{R}y_{126}, \label{MR}
\end{align}
where $M_U, M_D, M_E, M_{\nu_D}, M_{\nu_R}$ denote the mass matrices of up quarks, down quarks, charged leptons, Dirac neutrinos and Majorana right-handed neutrinos, respectively.
The properties of these Yukawa coupling matrices are given in Eq.~\eqref{eq:Yproperty}, which dictate that \(y_{10}\) and \(y_{126}\) are real symmetric, whereas \(y_{120}\) is real antisymmetric.
The VEVs of the doublet fields have been defined in Eq.~\eqref{eq:VEVs}. 
We will work in a basis where \(y_{126}\) is positive and diagonal. Furthermore, without loss of generality, \(v_{126}^u\) can be taken to be real. Moreover, by performing a gauge rotation in the \(U(1)_{B-L}\) part of the symmetry, one can also make \(v_R\) real. As noted in the previous section we also take $U_{41}$ to be real without loss of generality. Explicitly, we write
\begin{align}
&v_{10}=|v_{10}| e^{i \alpha}, \quad   v^u_{126}=|v^u_{126}|, \quad   v^d_{126}=|v^d_{126}| e^{i \gamma}, \quad v_{R}= |v_{R}|, 
\\
& v^{(1)}_{120}+v^{(15)}_{120} = |v^{(1)}_{120}+v^{(15)}_{120}| e^{i \beta}, \quad v^{(1)\ast}_{120}-3v^{(15)\ast}_{120}= |v^{(1)\ast}_{120}-3v^{(15)\ast}_{120}| e^{i \rho}.
\end{align}
With that, we define the following matrices and quantities:
\begin{align}
&S=|v^u_{126}|  y_{126}, \quad 
D=|v_{10}|  y_{10} , \quad 
A=|v^{(1)}_{120}+v^{(15)}_{120}| y_{120},
\\
&r_1=\frac{|v^d_{126}|}{|v^u_{126}|}e^{i\gamma},\;\; r_2=\frac{|v^{(1)\ast}_{120}-3v^{(15)\ast}_{120}|}{|v^{(1)}_{120}+v^{(15)}_{120}|} e^{i\rho},\;\; c_{R}=\frac{|v_{R}|}{|v^u_{126}|},\label{cR}
\end{align}
where \(D, S, A\) are real matrices,
which allows us to re-write the mass matrices as:
\begin{align}
&M_U= e^{i \alpha} D + S + e^{i \beta} A \equiv vy_U, \label{E1}\\
&M_D= e^{-i \alpha} D + r_1 S +e^{-i \beta} A \equiv vy_D,\\
&M_E= e^{-i \alpha} D -3 r_1 S + r_2 A \equiv vy_E, \label{ME} \\
&M_{\nu_D}= e^{i \alpha} D -3 S + r_2^* A \equiv vy_{\nu_D}, \label{MDnu}\\
&M_{\nu_{R}}=c_{R} S, \label{E2}
\end{align}
where $v\simeq 174$ GeV is the electroweak VEV and the Yukawa interactions between fermions and the SM Higgs doublet $h$ are defined as
	\begin{eqnarray}
	-{\cal L} & \supset & \overline{U}y_U Q\epsilon h +\overline{D} y_D Q h^{*} +\overline{E} y_E \ell h^{*} +\overline{N}y_{\nu_D} \ell\epsilon  h +{\rm h.c.},\label{eq:Yukawa_light_Higgs}
\end{eqnarray}
where we have suppressed the family indices.
Here $\epsilon$ denotes the $SU(2)_L$ totally antisymmetric tensor, $Q$ and $\ell$ denote $SU(2)_L$ doublets of quarks and leptons respectively, and, $U$, $D$ and $E$ denote $SU(2)_L$ singlets of up-type quarks, down-type quarks and charged leptons, respectively.

As shown in Ref.~\cite{Babu:2016bmy}, type-II seesaw dominance for light neutrino masses is not viable and hence we will consider only type-I seesaw dominance with light neutrino mass matrix given by
\begin{align}
m_\nu = - M^T_{\nu_D} M^{-1}_{\nu_R} M_{\nu_D} .  \label{Mass:nu}
\end{align}

\textbf{Parameter counting} --- At this point, it is important to count the number of parameters relevant for fermion masses and mixings. Since all Yukawa couplings are real due to $CP$ invariance, and in our chosen basis \(y_{126}\) is diagonal, we have 6 real parameters in \(y_{10}\), 3 real and positive entries in \(y_{126}\), and 3 real parameters in the antisymmetric coupling \(y_{120}\). The VEV ratios \(r_1\) and \(r_2\) are complex, whereas \(c_R\) is real. Furthermore, the mass matrices in Eqs.~\eqref{E1}--\eqref{E2} contain two additional phases, namely \(\alpha\) and \(\beta\), totaling 19 parameters -- 15 magnitudes and 4 phases.

It is interesting to point out that in the Peccei--Quinn (PQ)~\cite{Peccei:1977hh} symmetric version, \(SO(10) \times U(1)_\mathrm{PQ}\)~\cite{Babu:1992ia}, utilizing a complex $10_H$ and a complex $126_H$ there are also  19 parameters, specifically 12 magnitudes and 7 phases. However, as will be shown below, compared to the PQ version, the proposed model here can fit the observables somewhat better, with a lower $\chi^2$ than in the PQ version (see, for example, Ref.~\cite{Babu:2020tnf}, which obtains a total $\chi^2\sim 8$ for the scenario with PQ symmetry). We would also like to compare our model with those featuring spontaneous $CP$ violation. In the non-supersymmetric framework, the model proposed in Ref.~\cite{Patel:2022xxu} contains a larger number of parameters---specifically, 21 parameters in the Yukawa sector. Moreover, the Higgs potential in that scenario permits more than one fundamental complex parameters, in contrast to our setup. Regarding the supersymmetric model of spontaneous $CP$ violation in Ref.~\cite{Mimura:2019yfi} (for earlier works along this line of investigation, employing older experimental data, see Refs.~\cite{Dutta:2005ni,Grimus:2006rk}), it is constructed by introducing additional discrete symmetries and incorporating non-renormalizable terms in the Yukawa superpotential. The Yukawa sector of this model also requires 21 parameters, which include supersymmetric threshold corrections, in order to correctly reproduce the fermion spectrum.  

In summary, our model has 19 parameters in the Yukawa sector to fit 19 observables: 9 charged fermion masses, 3 CKM angles, 1 CKM phase, 2 neutrino mass-squared differences, 3 leptonic mixing angles, and 1 baryon asymmetry parameter. The CP-violating Dirac phase \(\delta_\mathrm{PMNS}\) in the neutrino sector (the 20th observable) has not been measured yet; therefore, we do not fit it, but will have this as a prediction of our framework. Similarly, the two currently unknown Majorana phase in the neutrino sector as well as the lightest neutrino mass, or equivalently the effective mass $m_{\beta\beta}$ relevant for neutrinoless double beta decay and the sum of the neutrino masses $\sum_i m_i$ relevant for cosmology will be constrained to be in narrow ranges.

%%%%%%%%%%%%%%%%%%%%%%%%%%%%%%%%%%%%%%%%%%%%%%%%%%%%%%%%%%%%%
\subsection{Numerical analysis and benchmark fit} 
In this section, we undertake a comprehensive fit of the fermion masses and mixings  arising from the mass matrices of  Eqs.~\eqref{E1}--\eqref{Mass:nu}, simultaneously incorporating the baryon asymmetry of the universe. Our fitting strategy begins with the use of approximate expressions for baryon asymmetry via leptogenesis, following the approach in Ref.~\cite{Blanchet:2011xq}, to quickly pinpoint promising parameter sets capable of generating adequate asymmetry. For reviews on leptogenesis, see for example Refs.  \cite{Buchmuller:2004nz,Nir:2007zq,Davidson:2008bu,Pilaftsis:2009pk,DiBari:2012fz,Fong:2012buy,Chun:2017spz,Dev:2017trv, Bodeker:2020ghk}. Once such candidate points are identified and a satisfactory fit is achieved, we perform a detailed numerical computation of the baryon asymmetry using the full formalism as detailed in our previous paper in Ref.~\cite{Babu:2024ahk}. This allows us to verify that the predicted baryon asymmetry matches the observed value
in terms of baryon-to-photon number density today $\eta_B = (6.12 \pm 0.04) \times 10^{-10}$~\cite{Planck:2018vyg}.
A summary of the fitting methodology is provided below.

The 19 free parameters listed above are randomly varied at the GUT scale, which we take as \(M_{\rm GUT} = 2 \times 10^{16}\) GeV. We then evolve the relevant Yukawa couplings defined in Eqs.~\eqref{E1}--\eqref{MDnu} and the right-handed neutrino mass matrix of Eq.~\eqref{E2} from \(M_{\rm GUT}\) down to the electroweak scale \(M_Z\), using the full SM plus Type-I seesaw renormalization group equations (RGEs). During this running, heavy right-handed neutrinos are sequentially integrated out at their respective mass thresholds. Our model implementation in the \texttt{REAP} package~\cite{Antusch:2005gp} handles this procedure efficiently. Since our model predicts an intermediate symmetry breaking scale $M_{\rm int} \sim 10^{14}$ GeV-
close to the GUT scale—it suffices for our analysis to consider only the SM+Type-I seesaw RGEs below the GUT scale.

\FloatBarrier
\begin{table}[t!]
\centering
\footnotesize
\resizebox{0.9\textwidth}{!}{
\begin{tabular}{|c|c|c|c|}
\hline
\textbf{Observables} & \multicolumn{3}{c|}{\textbf{Values at $M_Z$ scale}} 
 \\ 

\cline{2-4}
 &\textbf{Experimental Input}&\textbf{Benchmark Fit Value}&\textbf{Pull} \\
\hline\hline

\rowcolor{teal!20}$y_u/10^{-6}$       & $6.65 \pm 2.25$     & 6.230 & -0.186 \\ \hline 
\rowcolor{teal!20}$y_c/10^{-3}$       & $3.60 \pm 0.11$     & 3.599 & -0.004 \\ \hline
\rowcolor{teal!20}$y_t$               & $0.98605 \pm 0.00865$ & 0.986 & -0.05 \\ \hline\hline

\rowcolor{red!13}$y_d/10^{-5}$       & $1.645 \pm 0.165$   & 1.646 & 0.009 \\ \hline
\rowcolor{red!13}$y_s/10^{-4}$       & $3.125 \pm 0.165$   & 3.122 & -0.014 \\ \hline
\rowcolor{red!13}$y_b/10^{-2}$       & $1.639 \pm 0.015$   & 1.639 & 0.011 \\ \hline\hline

\rowcolor{blue!10}$y_e/10^{-6}$       & $2.79474 \pm 0.02794$ & 2.794 & 0.0027 \\ \hline
\rowcolor{blue!10}$y_\mu/10^{-4}$     & $5.89986 \pm 0.05899$ & 5.900 & 0.0041 \\ \hline
\rowcolor{blue!10}$y_\tau/10^{-2}$    & $1.00295 \pm 0.01002$ & 1.003 & 0.0073 \\ \hline\hline

\rowcolor{yellow!13}$\theta^\mathrm{CKM}_{12}/10^{-2}$ & $22.735 \pm 0.072$ & 22.734 & -0.0036 \\ \hline
\rowcolor{yellow!13}$\theta^\mathrm{CKM}_{23}/10^{-2}$ & 4.208 $\pm$ 0.064 & 4.208 & 0.0012 \\ \hline
\rowcolor{yellow!13}$\theta^\mathrm{CKM}_{13}/10^{-2}$ & 0.364 $\pm$ 0.013 & 0.364 & 0.0058 \\ \hline
\rowcolor{yellow!13}$\delta_\mathrm{CKM}$ (deg) & $69.21 \pm 3.09$ & 69.21 & -0.0009 \\ \hline\hline

\rowcolor{green!10}$\Delta m^2_{21}/10^{-5} (\mathrm{eV}^2)$ & 7.49 $\pm$ 0.19 & 7.489 & -0.003 \\ \hline
\rowcolor{green!10}$\Delta m^2_{31}/10^{-3} (\mathrm{eV}^2)$ & 2.5345 $\pm$ 0.024 & 2.534 & 0.0042 \\ \hline\hline

\rowcolor{orange!13}$\sin^2 \theta_{12}$ & 0.3075 $\pm$ 0.0115 & 0.3075 & 0.0067 \\ \hline
\rowcolor{orange!13}$\sin^2 \theta_{23}$ & 0.5596 $\pm$ 0.0135 ($\dagger$) & 0.4653 & -1.23 \\ \hline
\rowcolor{orange!13}$\sin^2 \theta_{13}$ & 0.02193 $\pm$ 0.00056 & 0.02192 & -0.0013 \\ \hline\hline

\rowcolor{cyan!40}$\eta_B/10^{-10}$& $6.12\pm 0.04$ &8.2& $(*)$ \\\hline\hline

\hline
$\chi^2$&-&-&1.57 \\
\hline

\end{tabular}

}
\caption{Benchmark fit values of the observables along with their experimentally measured values. $(*)$Since the baryon asymmetry parameter \(\eta_B\) is computed using an analytical approximation during the fitting procedure, but the full numerical result is presented in this table, the corresponding pull is not meaningful and is therefore not quoted. ($\dagger$)It is important to note that the neutrino mixing angle \(\theta_{23}\) is currently allowed to lie in either the first or second octant~\cite{NUFIT}. In this Table, we quote the best-fit value obtained from the global fit to neutrino oscillation data. However, during the fitting procedure, we allow \(\theta_{23}\) to vary over its entire experimentally viable range.
  }\label{result}
\end{table}

Our fitting procedure involves performing a \(\chi^2\) minimization at the scale \(M_Z\), where
\begin{align}
\chi^2 = \sum_k \left(\frac{T_k - O_k}{E_k}\right)^2,    
\end{align}
with \(T_k\), \(O_k\), and \(E_k\) representing the theoretical prediction, experimental central value, and \(1\sigma\) uncertainty of the \(k\)-th observable, respectively. The experimental input values for charged and neutral fermion sectors are taken from Refs.~\cite{Antusch:2013jca} and~\cite{NUFIT,Esteban:2020cvm}. In addition to the fermion masses and mixing parameters, the \(\chi^2\) function also incorporates the baryon asymmetry of the universe, \(\eta_B\) (see discussion below). Since this asymmetry arises from the decays of the next-to-lightest right-handed neutrino \(N_2\), \(\eta_B\) is computed using the Dirac Yukawa coupling \(Y_{\nu_D}\) and the right-handed neutrino masses \(M_{\nu_R}\) evaluated at the scale \(M_2\), corresponding to the mass of \(N_2\). 

Our best-fit benchmark results are presented in Table~\ref{result}, which also lists the experimental input values of the observables. The model parameters corresponding to this benchmark fit are summarized in Appendix~\ref{app:bestfit}. Finally, predictions of some of the physical quantities obtained from the benchmark best fit are summarized in Table~\ref{tab:example}.

\begin{table}[t]
\centering
\footnotesize
\resizebox{0.99\textwidth}{!}{
\begin{tabular}{|c|c|c|c|}
\hline
\textbf{Quantity} &  \textbf{Exp./bounds}  &  \textbf{Benchmark prediction} &  \textbf{ $2\sigma$ HPD } \\
\hline \hline

\rowcolor{green!10}$m_1$ (meV) &-   & 0.09   & (0.071, 0.127)  \\ \hline
\rowcolor{green!10}$\sum_i m_i$ (meV)  & <($87$\textendash$120$)~\cite{ParticleDataGroup:2024cfk} &  59.09  &  (58.52, 59.65) \\ \hline
\rowcolor{green!10}$m_{\beta\beta}$ (meV)  &<($28$\textendash$122$)~\cite{KamLAND-Zen:2024eml} & 1.8  & (1.20, 2.29)  \\ \hline \hline

\rowcolor{orange!13}$\delta_\mathrm{PMNS}$ (deg)  & (96-422) $3\sigma$~\cite{NUFIT} & 331.014   & (-37.8, 31.1) \\ \hline \hline

\rowcolor{cyan!10}$\left(M_1, M_2, M_3\right)$ (GeV)    &- & $\left( 8.99\cdot 10^4, 1.59\cdot 10^{12}, 4.11\cdot 10^{14} \right)$  & -% (8.07, 10.9), (1.33, 1.72), (3.82, 4.86)  
\\
\hline
\end{tabular}
}
\caption{Derived physical quantities from the benchmark best fit. Here $m_{\beta\beta}$ is the neutrinoless double beta decay parameter. For some of these quantities, available experimental bounds are listed in the second column. Moreover, the fourth column gives the $2\sigma$ highest posterior Density
(HPD) of the MCMC analysis (see Sec.~\ref{MCMC}).  }
\label{tab:example}
\end{table}

A few remarks regarding the fit are in order.
\begin{itemize}
\item As shown in Table~\ref{result}, all fermion masses and mixing parameters are reproduced with excellent accuracy when compared to their experimentally measured values. The total \(\chi^2\) for this best-fit point is 1.57, with the dominant contribution coming from the observable \(\theta_{23}\) in the neutrino sector. This arises because our analysis shows that, when the baryon asymmetry requirement is imposed, the model favors \(\theta_{23} < 45^\circ\). A value of \(\theta_{23}\) in the first octant leads to \(\Delta \chi^2_{\theta_{23}} \sim 1.5\)~\cite{NUFIT}. We have verified that, if the baryon asymmetry constraint is relaxed, solutions with \(\theta_{23} > 45^\circ\) can also be accommodated, resulting in a slightly smaller total \(\chi^2 \sim 0.7\).

\item Within this framework of spontaneous CP violation, our model predicts a normal mass ordering for the neutrinos.

\item A scenario with pure type-II seesaw dominance does not yield a viable fit to the data~\cite{Babu:2016bmy}.

\item Unlike Ref.~\cite{Babu:2024ahk}, when $CP$ symmetry is imposed in the theory, only the normal mass ordering for neutrinos is allowed.  

\item As mentioned above, for numerical efficiency, we use an analytical approximate solution for the baryon asymmetry parameter \(\eta_B\) during the fitting procedure. Therefore, we allow a larger uncertainty compared to the experimental error, which is sufficient for our purpose of first obtaining the correct sign and the right order of magnitude for $\eta_B$.   Once the best fit is obtained, we verify the result by solving the full numerical Boltzmann equations, with the corresponding value quoted in Table~\ref{result}. Since the analytical approximation and the numerical computation yield slightly different outcomes, the corresponding pull for \(\eta_B\)  is not meaningful and is therefore omitted in Table~\ref{result}.

\end{itemize}

%%%%%%%%%%%%%%%%%%
\begin{figure}[t]
\centering
\includegraphics[width=0.47\textwidth]{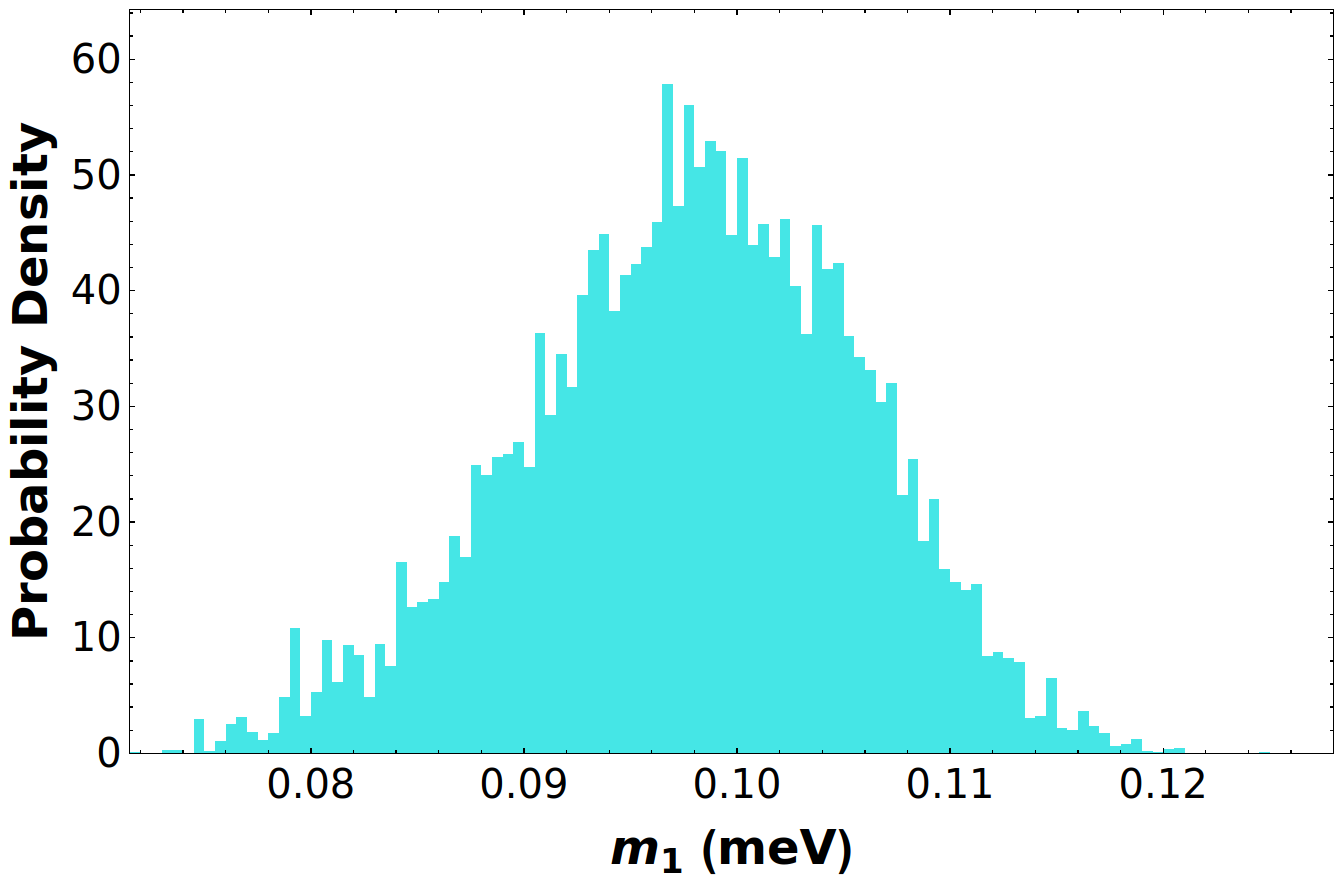}
\includegraphics[width=0.47\textwidth]{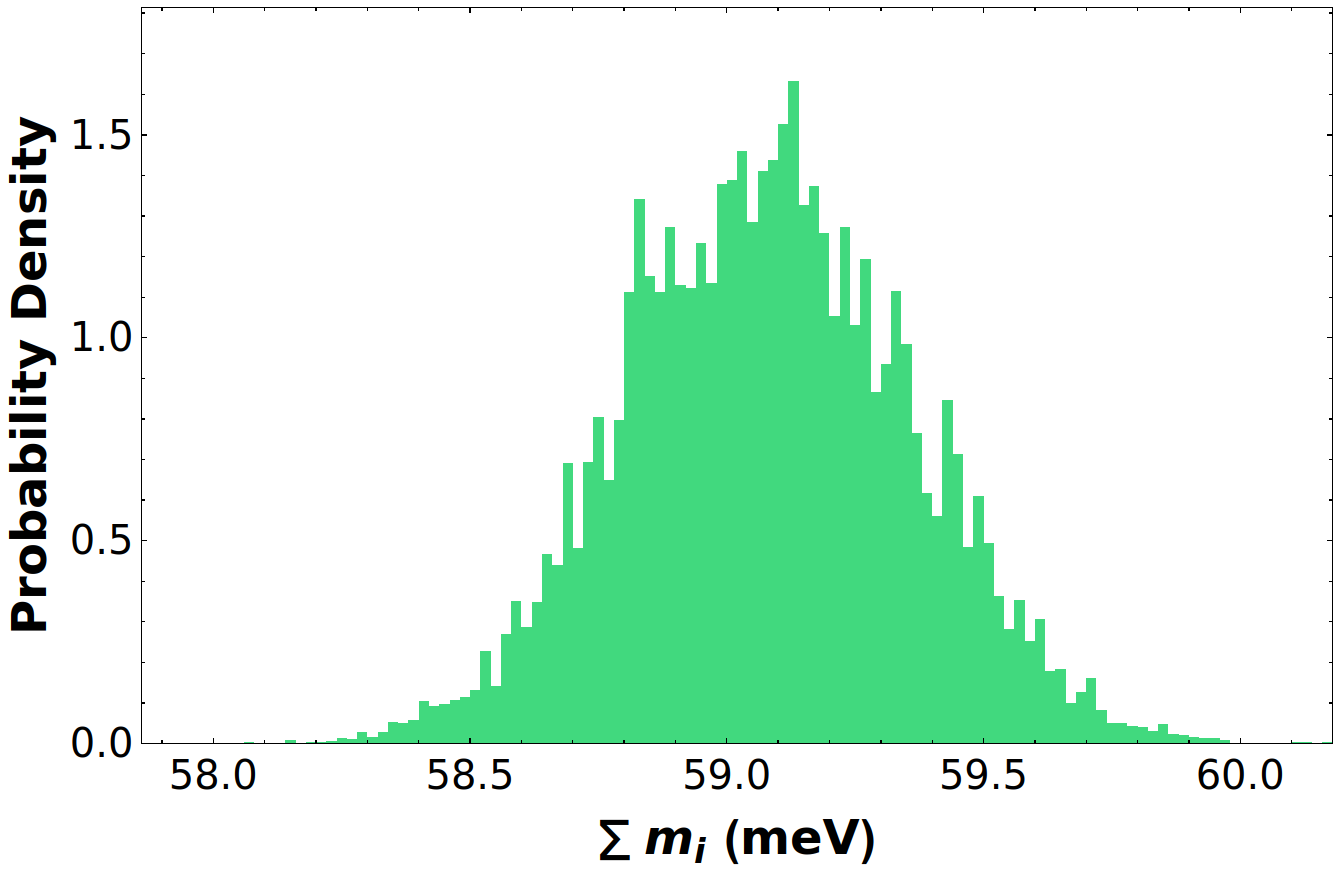}
\includegraphics[width=0.47\textwidth]{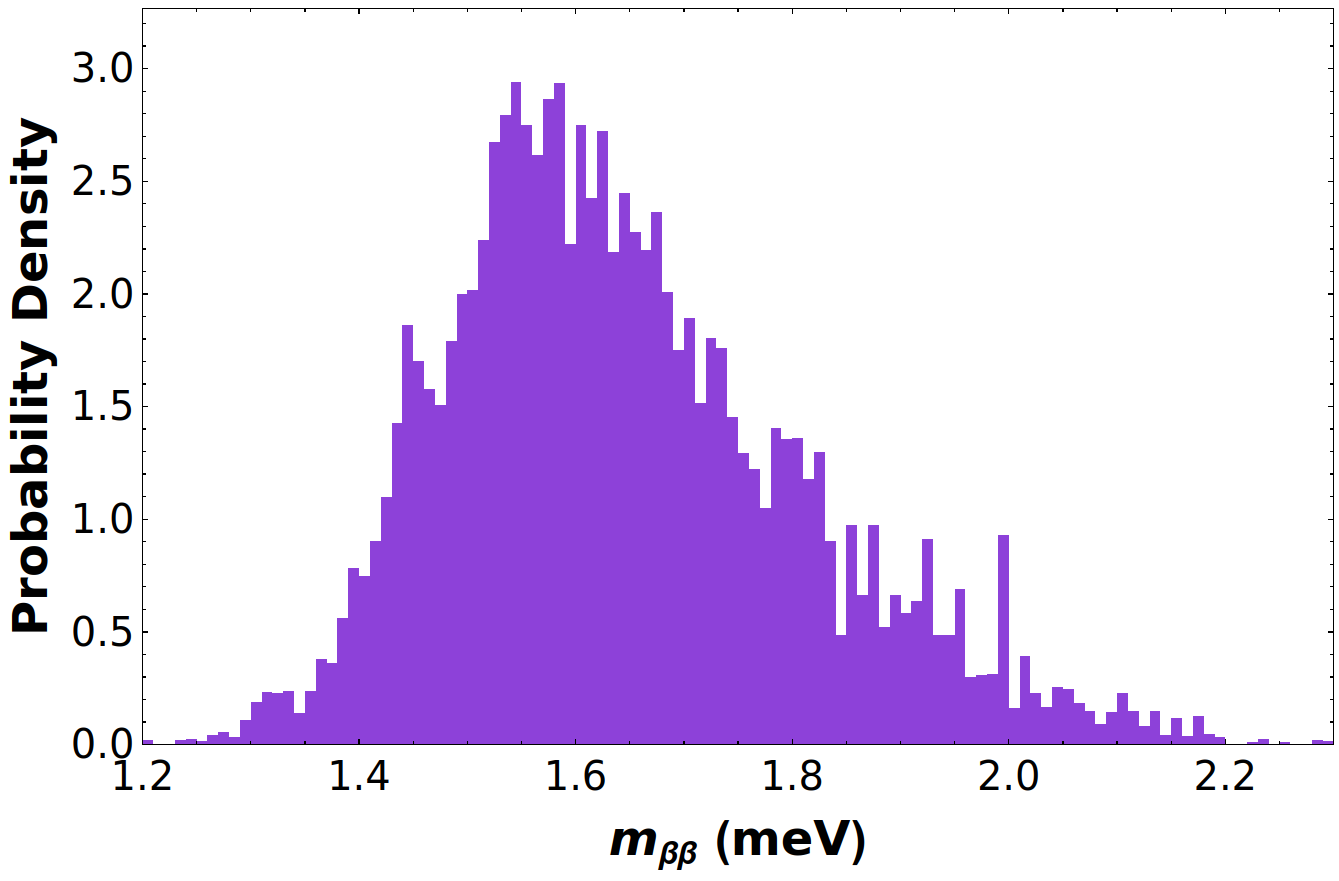}
\\
\includegraphics[width=0.47\textwidth]{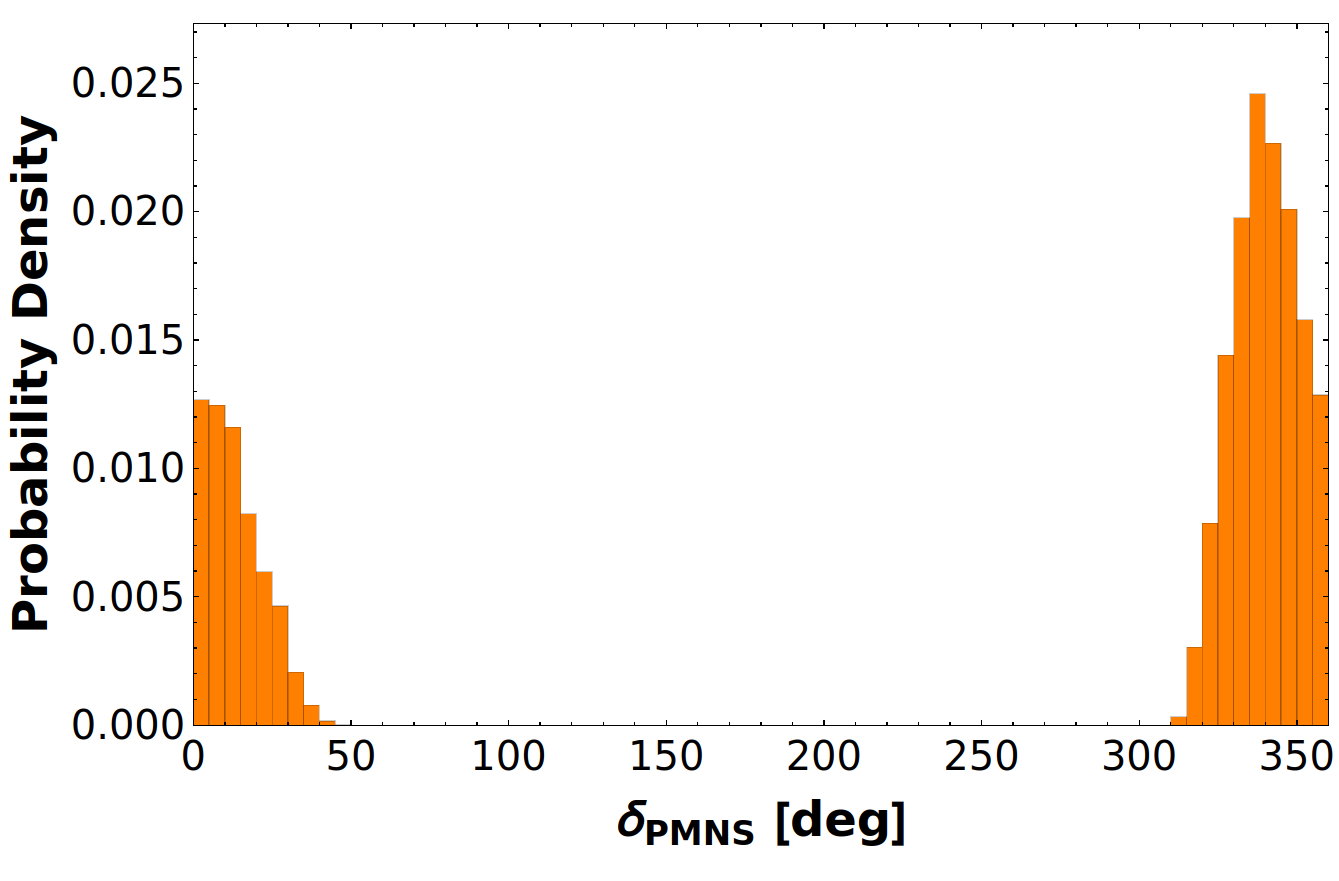}
\includegraphics[width=0.47\textwidth]{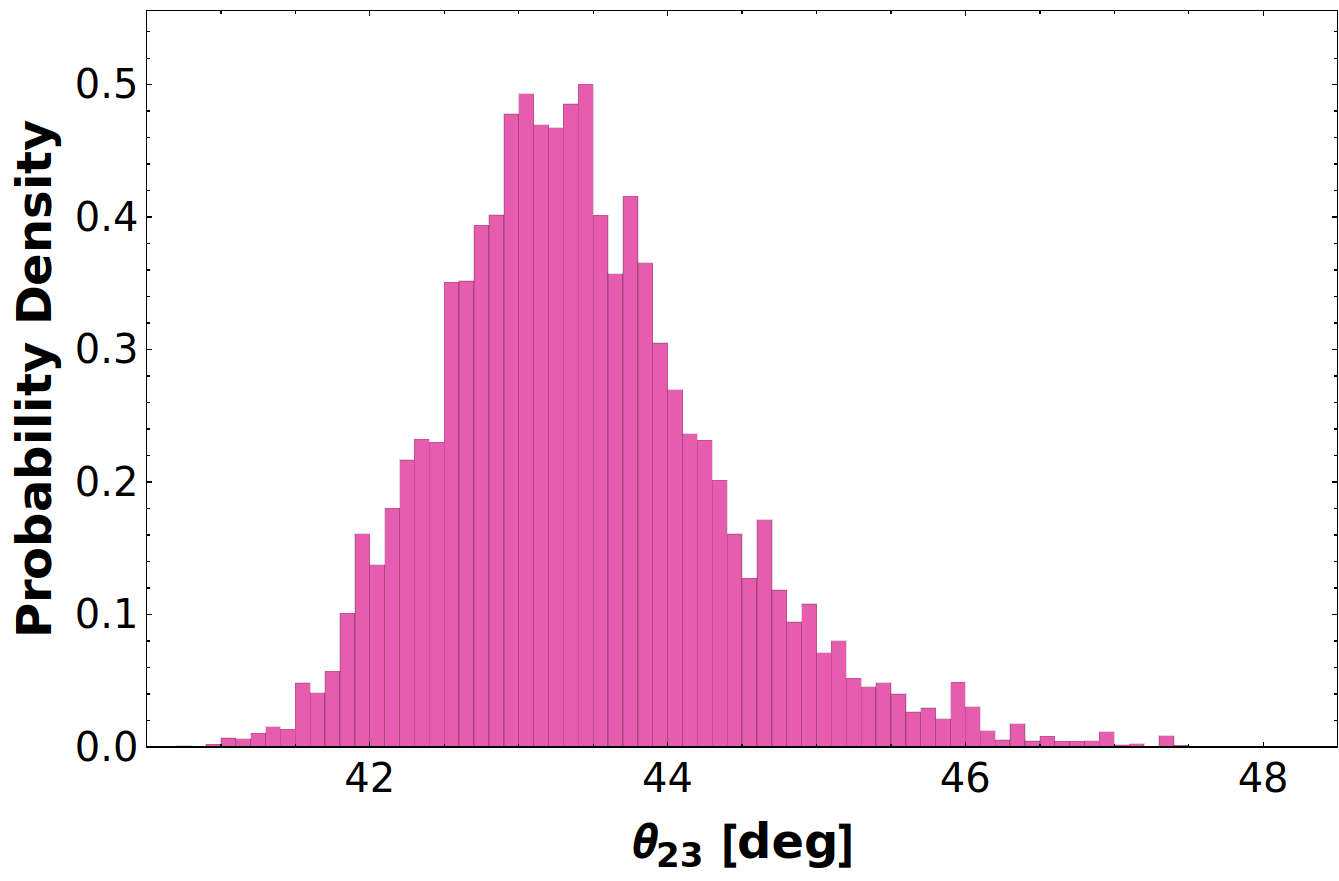}
\includegraphics[width=0.47\textwidth]{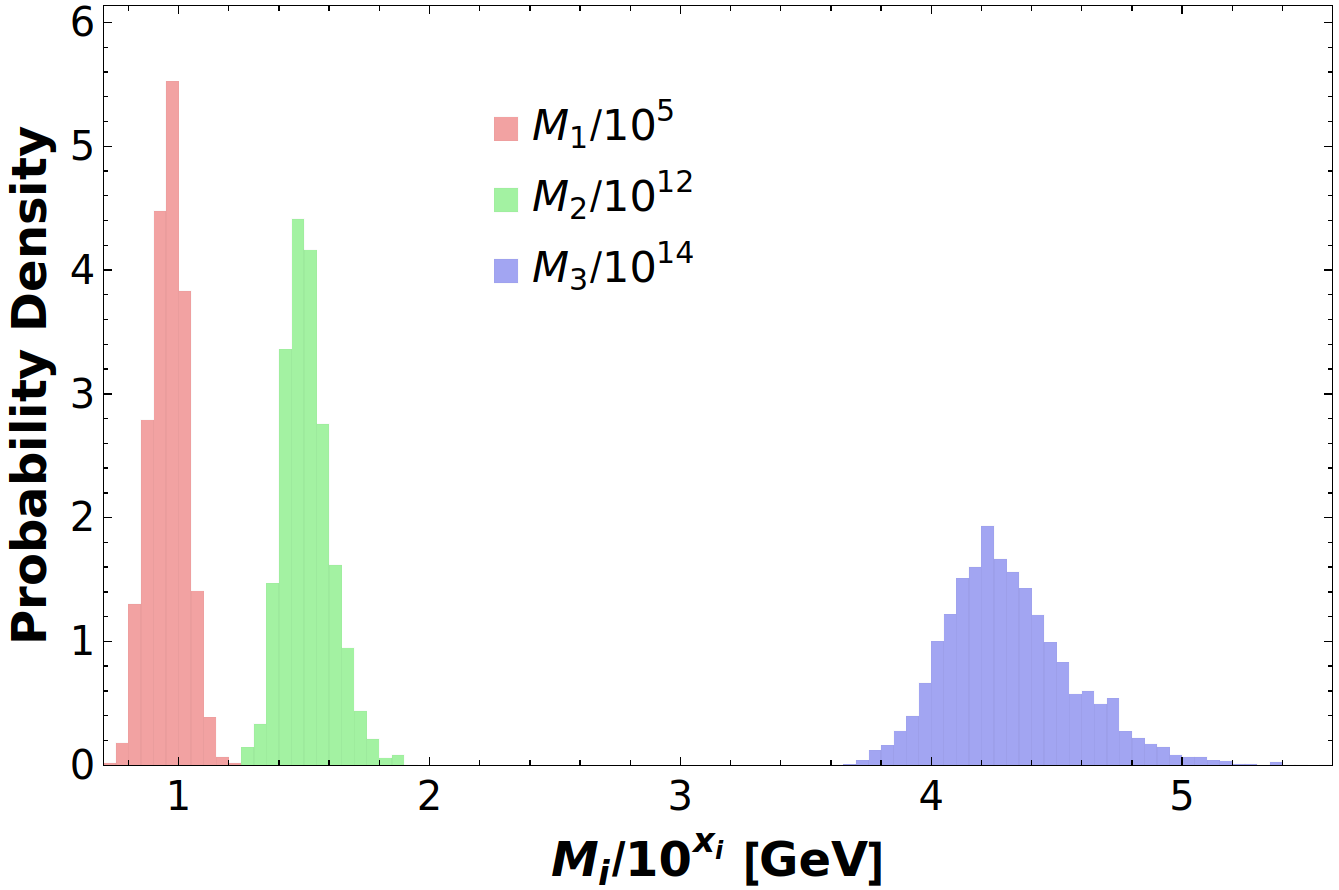}
\includegraphics[width=0.47\textwidth]{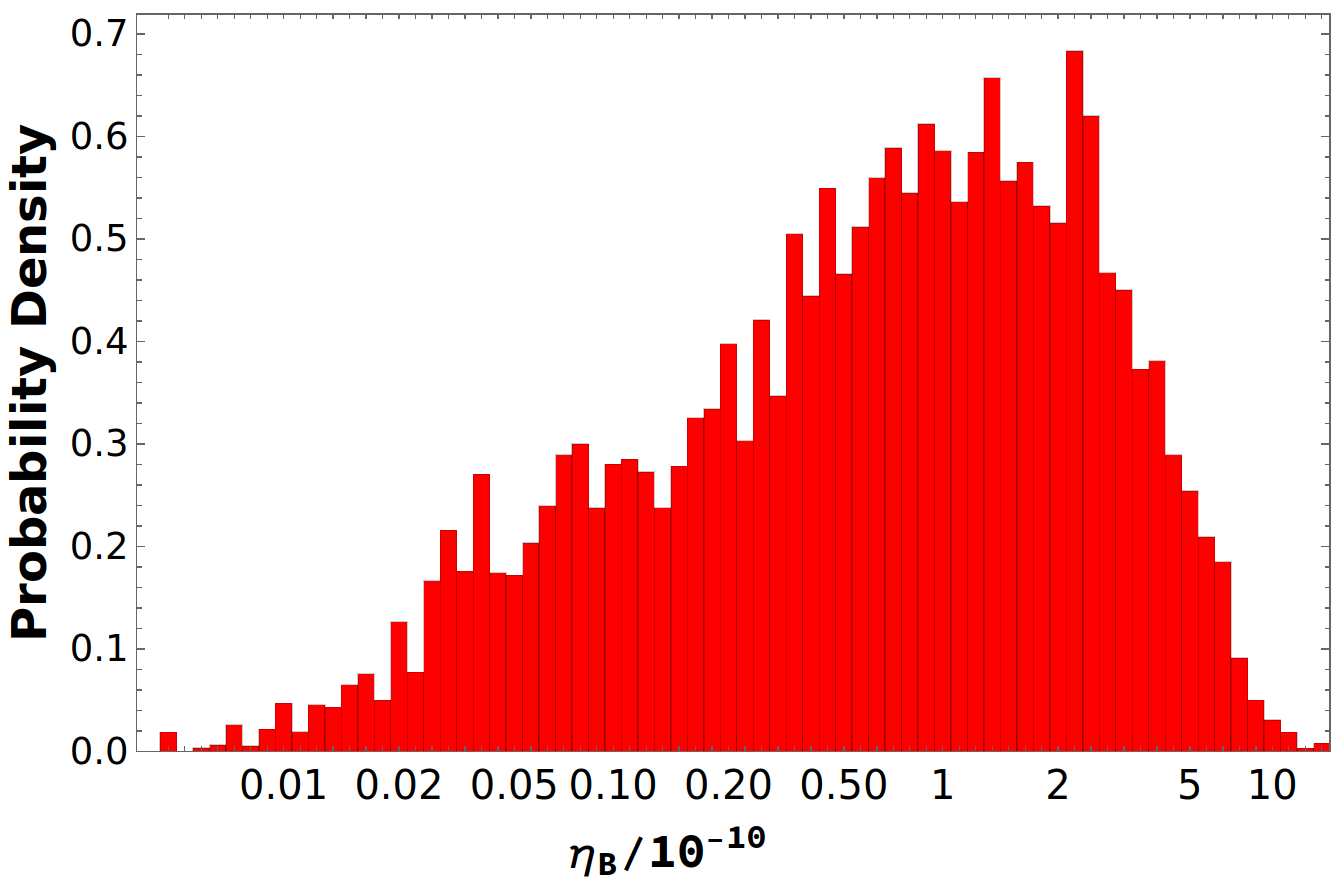}
\caption{ Likelihood region predicted from the MCMC analysis. See text for details. } \label{fig:MCMC}
\end{figure}
%%%%%%%%%%%%%%%%%%

Once a fit to the fermion masses is obtained, the flavor structure becomes fixed. Consequently, the proton decay branching ratios can be predicted. In our model, the two dominant decay channels are expected to be $p \to \overline{\nu}\pi^+$ and $p \to e^+\pi^0$~\cite{Babu:2016bmy}. 
 The lifetime for these decays have large uncertainties arising from threshold corrections which can only be estimated. Typically, these decays are within reach of SuperK and HyperK  with the lifetime lying in the $10^{33}-10^{35}$ yrs~\cite{Babu:2016bmy}.
Each of these channels is predicted to have a branching ratio close to $50\%$, while all other modes are subdominant. Therefore, measuring these branching ratios could provide a direct probe of the model.

Finally, we note that the baryon asymmetry parameter is sensitive to initial conditions since $N_2$ is in the weak washout regime while for $N_1$ washout, there is a protected direction in flavor space of $\ell$. Therefore, if there is an additional source generating the baryon asymmetry from the decay of particles that survive inflation, the value of $\eta_B$ will be modified. The result quoted in Table~\ref{result} assumes that baryon asymmetry is produced solely from leptogenesis assuming initial zero $N_i$ abundance. Assuming thermal initial abundance of $N_i$, which can be realized if the reheating temperature after inflation is much above $M_2$, we obtain an enhancement factor of 15.

%%%%%%%%%%%%%%%%%%%%%%%%%%%%%%%%%%%%%%%%%%%%%%%%%%%%%%%%%%%%%
\subsection{MCMC analysis} \label{MCMC}
Since the Yukawa sector spans a multi-dimensional parameter space, in this section we perform a Markov Chain Monte Carlo (MCMC)~\cite{Metropolis:1953am,Hastings:1970aa} analysis to study the probability distributions of several observables of particular interest. Our MCMC analysis begins from the best-fit solution obtained above and explores the likelihood regions of all observables as well as all relevant parameters involved in the fermion mass analysis. The MCMC results presented below comprise points with a total  $\chi^2$ value of approximately $\chi^2/n_{\mathrm{obs}} \simeq 2.4$, where $n_{\mathrm{obs}} = 19$ denotes the number of observables included in our fitting procedure to determine the best fit point. It is important to note that the baryogenesis constraint was omitted from the $\chi^2$ function during this MCMC analysis and we have utilized an analytical and approximate formula for $\eta_B$ in producing the right plot in the last row of Fig.~\ref{fig:MCMC}.

\begin{figure}[th!]
\centering
\includegraphics[width=0.31\textwidth]{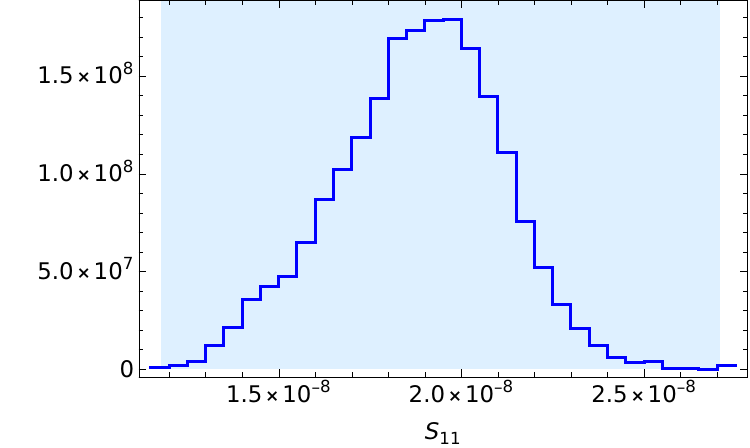}
\includegraphics[width=0.28\textwidth]{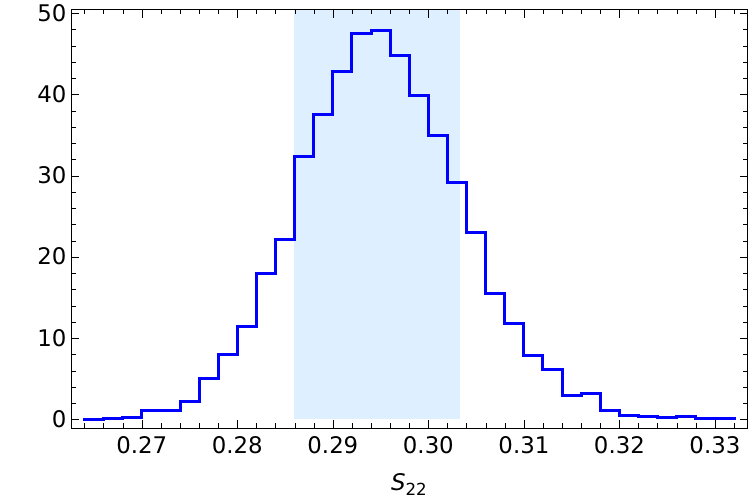}
\includegraphics[width=0.28\textwidth]{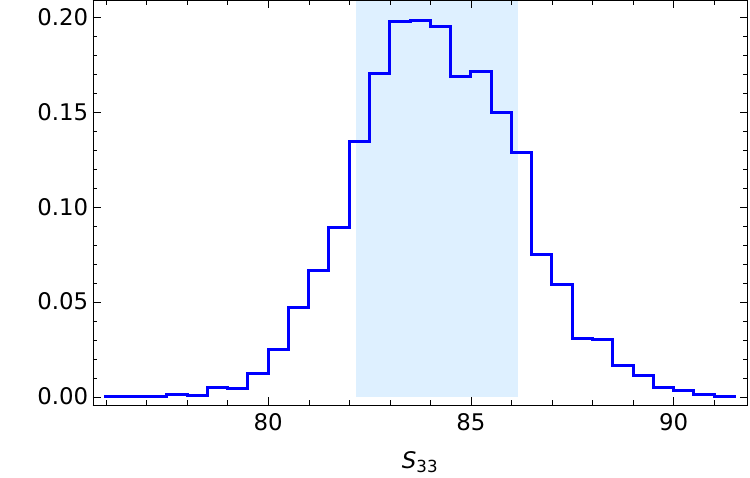}
\\

\includegraphics[width=0.28\textwidth]{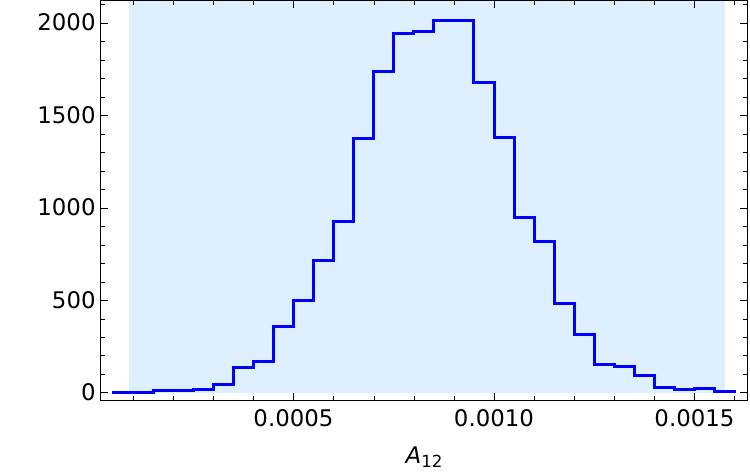}
\includegraphics[width=0.28\textwidth]{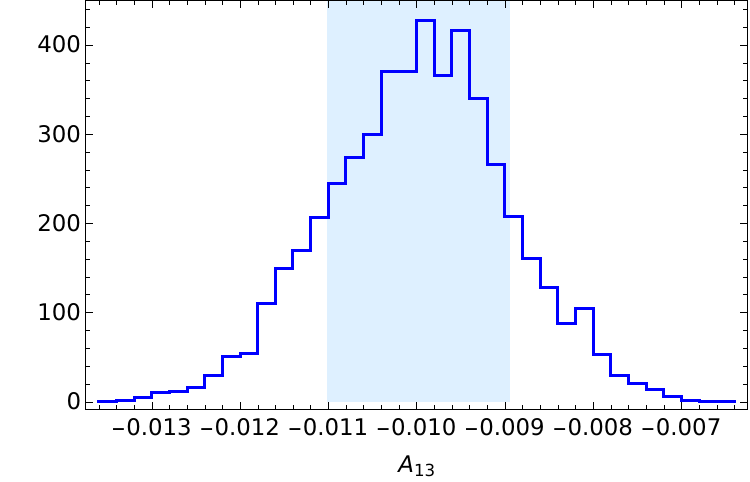}
\includegraphics[width=0.28\textwidth]{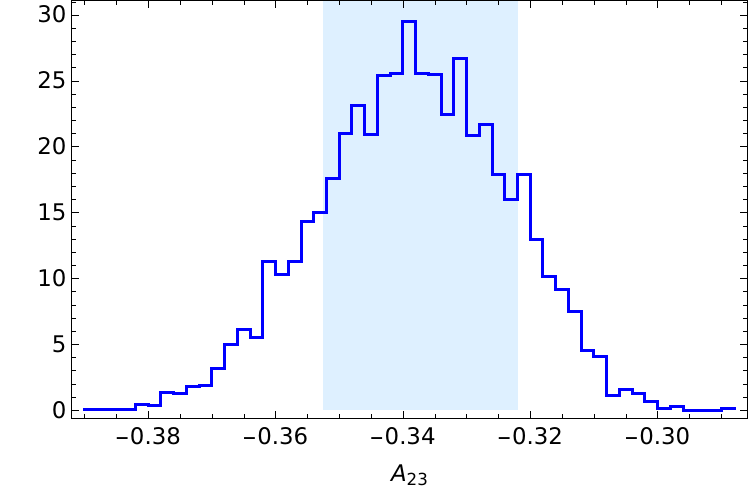}
\\

\includegraphics[width=0.28\textwidth]{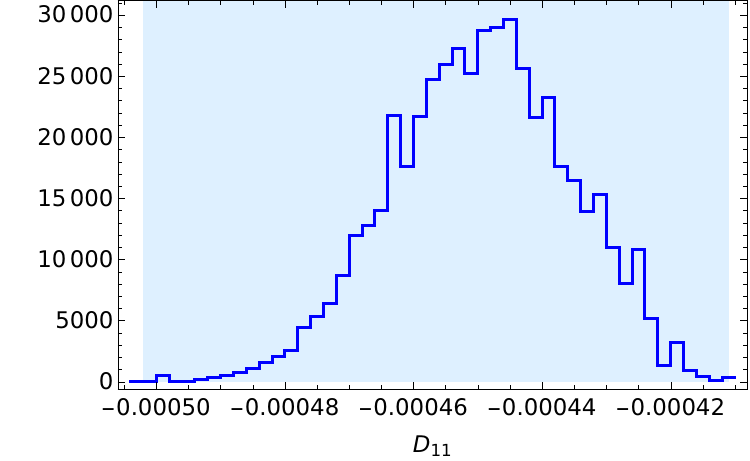}
\includegraphics[width=0.28\textwidth]{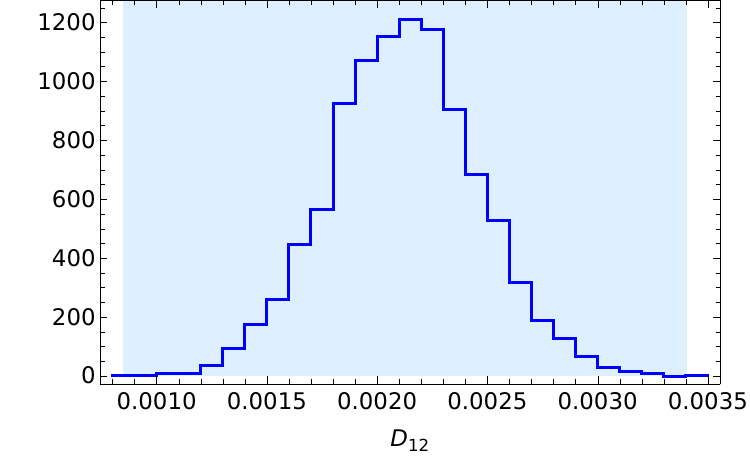}
\includegraphics[width=0.28\textwidth]{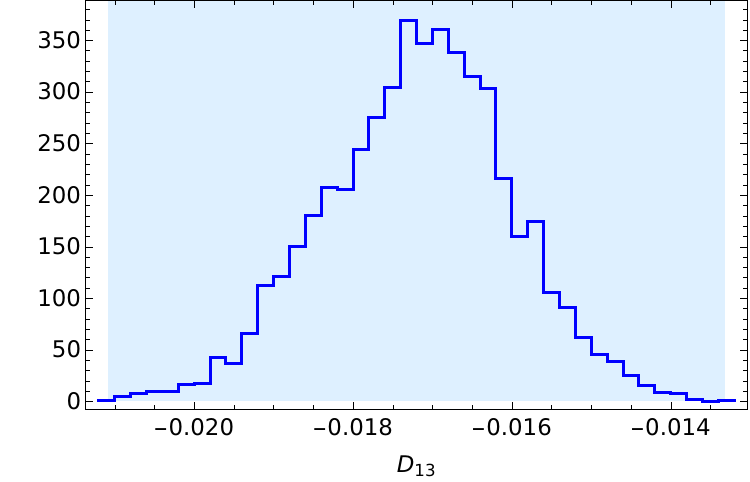}
\\

\includegraphics[width=0.28\textwidth]{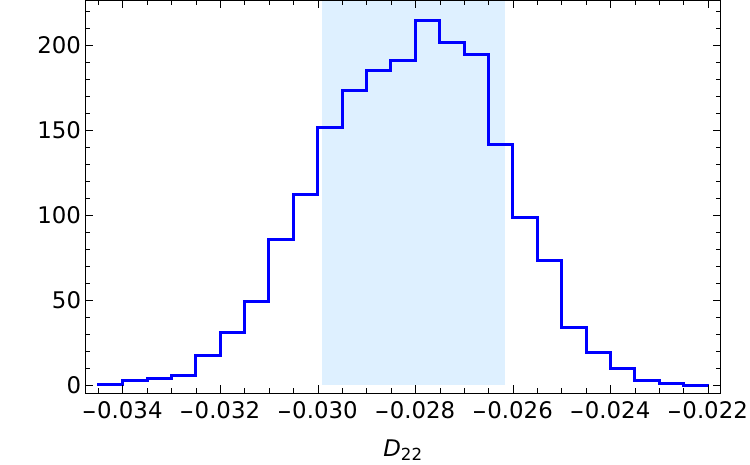}
\includegraphics[width=0.28\textwidth]{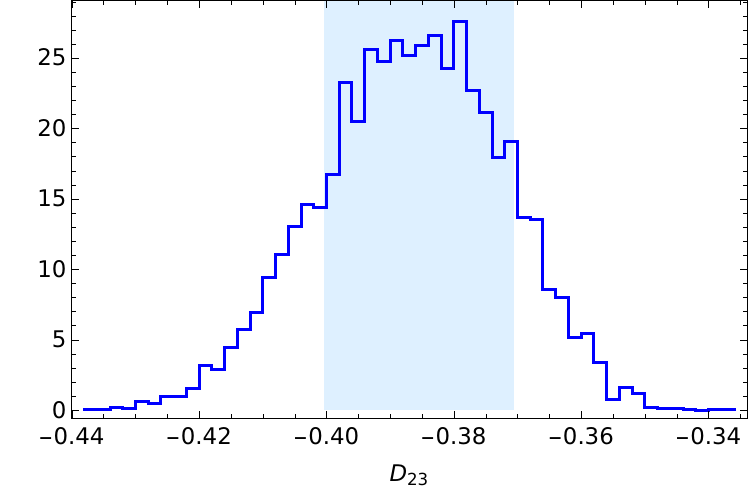}
\includegraphics[width=0.28\textwidth]{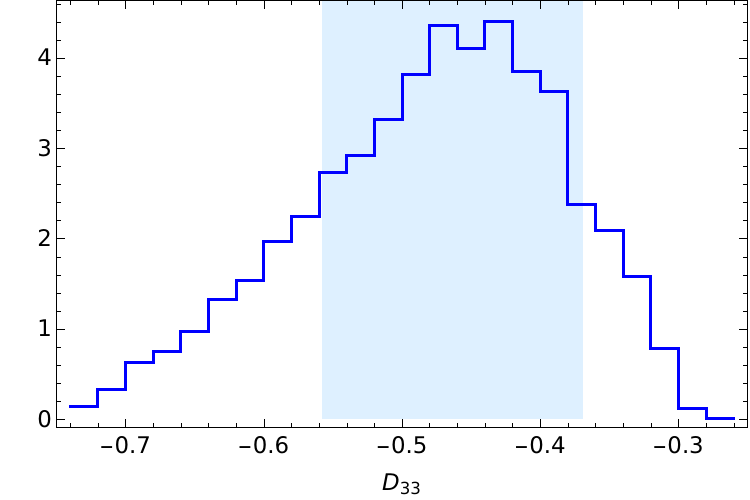}
\\

\includegraphics[width=0.28\textwidth]{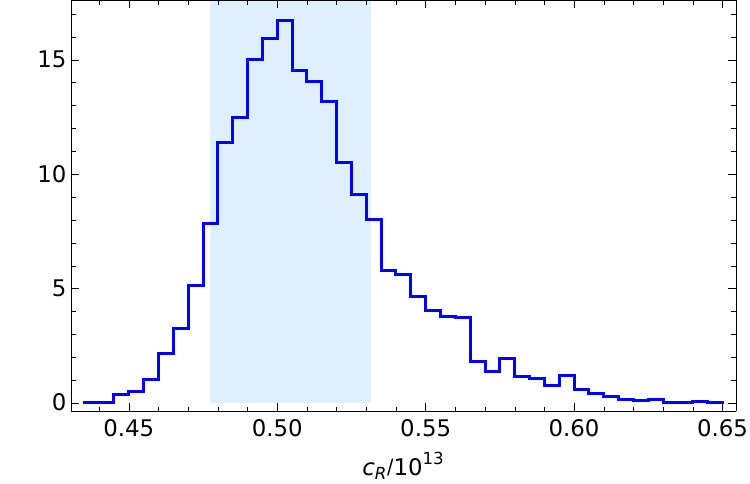}
\includegraphics[width=0.28\textwidth]{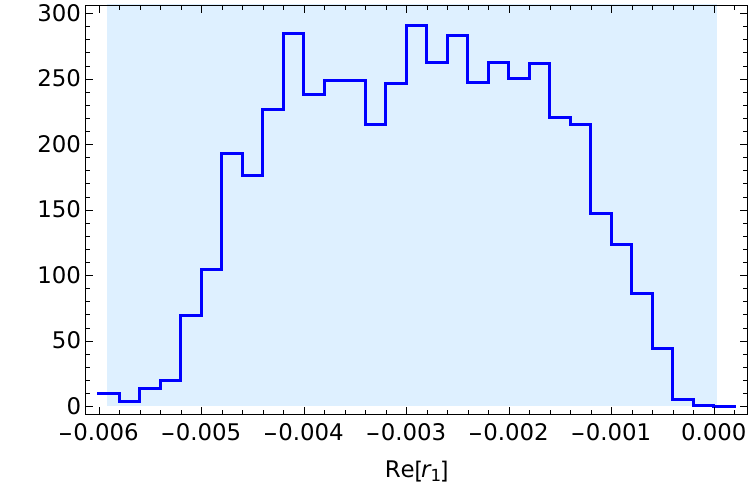}
\includegraphics[width=0.28\textwidth]{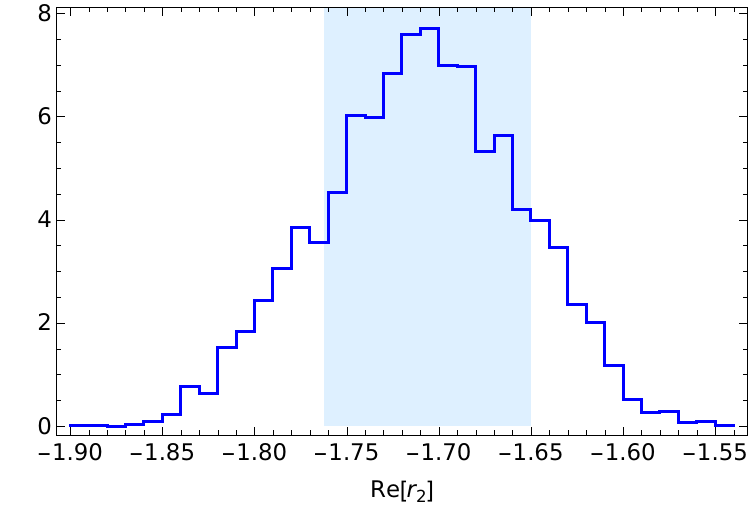}
\\

\includegraphics[width=0.28\textwidth]{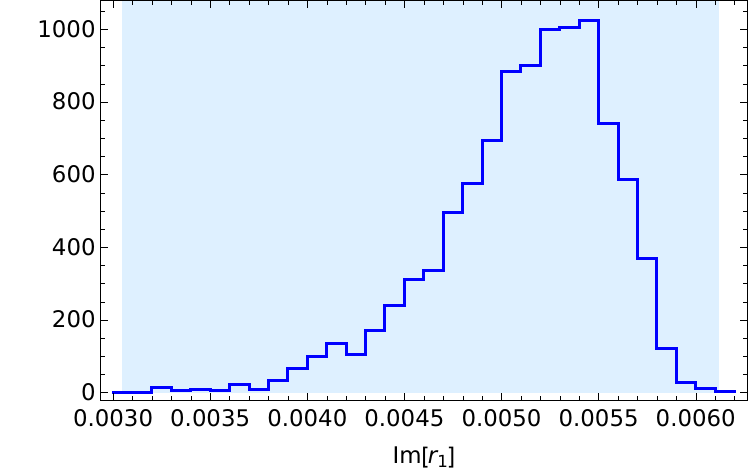}
\includegraphics[width=0.28\textwidth]{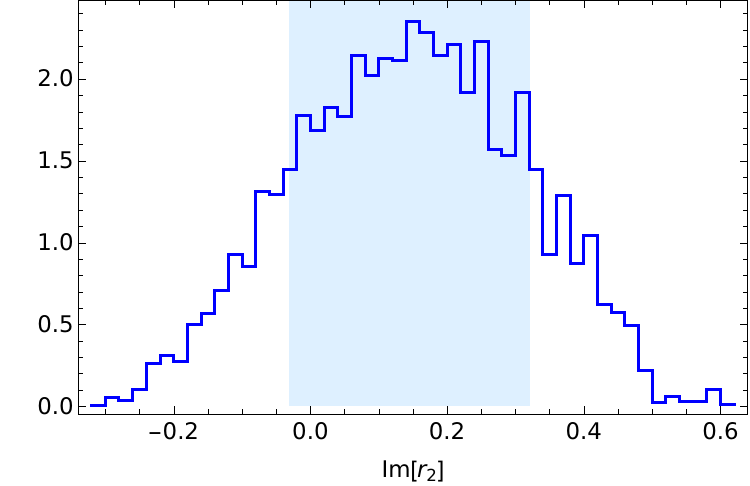}
\includegraphics[width=0.28\textwidth]{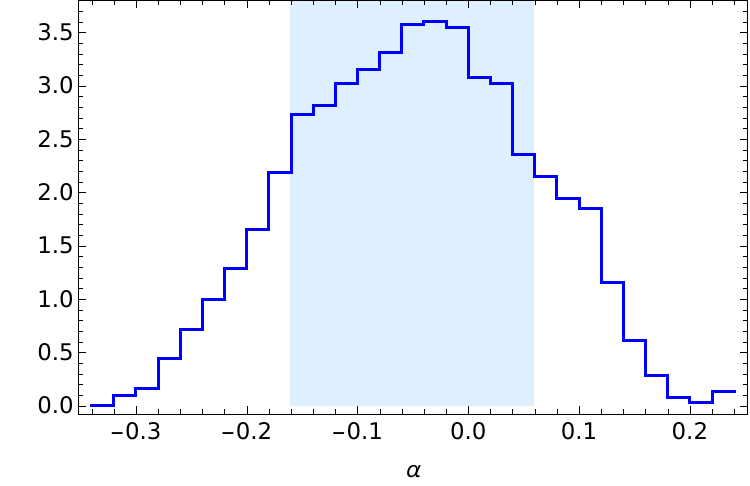}
\\

\includegraphics[width=0.28\textwidth]{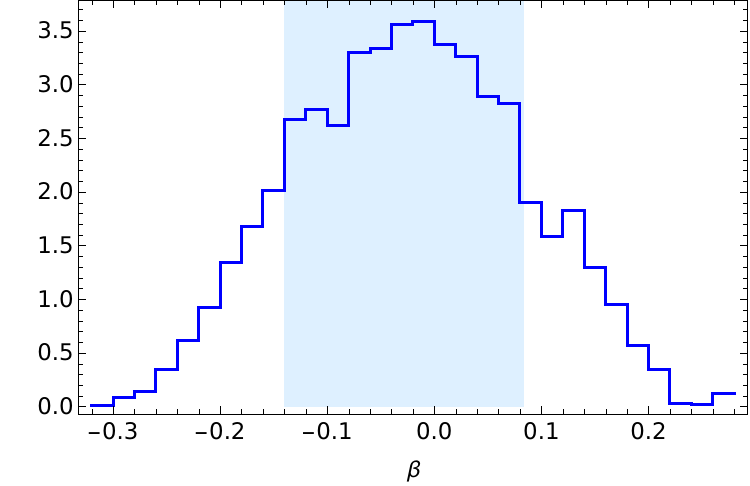}

\caption{ Probability distributions of the  parameters in the Yukawa sector  from the MCMC analysis. Shaded blue regions correspond to $2\sigma$ HPD intervals.     } \label{fig:PD}
\end{figure}

The MCMC analysis displayed in Fig.~\ref{fig:MCMC} indicates that the model favors the atmospheric mixing angle $\theta_{23}$ to reside predominantly in the first octant with a $2\sigma$ Highest Posterior Density (HPD) interval lies in the range $(41.5^\circ, 45.4^\circ)$. Additionally, the leptonic Dirac CP phase $\delta_{\mathrm{PMNS}}$ is constrained within the $2\sigma$  HPD  interval $(-38^\circ,\, +31^\circ)$. The lightest neutrino mass is found to be exceedingly small, approximately $m_1 \sim (0.071, 0.127)~\mathrm{meV}$, while the effective mass parameter relevant for neutrinoless double beta decay is predicted to be on the order of $m_{\beta\beta} \sim (1.20, 2.29)~\mathrm{meV}$. Our fitting procedure reveals a strongly hierarchical mass spectrum for the three right-handed neutrinos, given by $(M_1,\,M_2,\,M_3) \sim \left(10^{5},\, 10^{12},\, 5 \times 10^{14}\right)~\mathrm{GeV}$, which supports a successful \emph{$N_2$-dominated} leptogenesis scenario. For completeness, the probability distributions of the model parameters are depicted in Fig.~\ref{fig:PD}.

%%%%%%%%%%%%%%%%%%%%%%%%%%%%%%%%%%%%%%%%%%%%%%%%%%
%%%%%%%%%%%%%%%%%%%%%%%%%%%%%%%%%%%%%%%%%%%%%%%%%%
\section{Conclusions}\label{sec:conclusion}
In this work, we have developed a renormalizable $SO(10)$ grand unified theory featuring spontaneous $CP$ violation with a minimal Yukawa sector comprising only 19 real parameters, which provides an excellent fit to 19 observables: the six quark masses, three CKM mixing angles and the CKM phase; the three charged-lepton masses; the two neutrino mass-squared differences; the three PMNS mixing angles; and the baryon asymmetry parameter. Since $CP$ is spontaneously broken, all Yukawa couplings are real at the Lagrangian level, and all $CP$-violating phases — including the Dirac phases in the quark and leptonic sectors, the two Majorana phases in the neutrino sector, as well as those relevant for leptogenesis — arise solely from complex vacuum expectation values. This is achieved through a minimal Higgs sector featuring a $\mathbf{54}_H$ representation that is odd under $CP$, which simultaneously breaks the $SO(10)$ gauge symmetry and $CP$ symmetry spontaneously.  A remarkable feature of our model is that, owing to the minimality of the Higgs sector, a single high-scale phase is responsible for generating all low-energy $CP$ phases. This single phase appears in different combinations with other real parameters in the fermion mass matrices, which are represented by four effective  phase parameters in the fit.  

Our analysis predicts a normal ordering for neutrino masses, with the atmospheric mixing angle $\theta_{23}$ likely residing in the first octant and the leptonic Dirac phase $\delta_{\mathrm{PMNS}}$ favored within the range $(-38^\circ, +31^\circ)$. These predictions are expected to be tested in upcoming experiments, including DUNE~\cite{DUNE:2020jqi}, JUNO~\cite{JUNO:2015zny}  and Hyper-Kamiokande~\cite{Hyper-Kamiokande:2018ofw}. The lightest neutrino mass is predicted to be exceedingly small, $m_1 \sim \mathcal{O}(0.1)~\mathrm{meV}$, consistent with cosmological bounds~\cite{Shao:2024mag}, while the effective neutrinoless double beta decay parameter $m_{\beta\beta}$ is of order $1$ meV, which will pose a challenge for its discovery in generation experiments such as nEXO~\cite{nEXO:2021ujk}. The strongly hierarchical right-handed neutrino mass spectrum, spanning from $10^{5}$ to $5 \times 10^{14}$~GeV, naturally facilitates successful $N_2$-dominated leptogenesis consistent with cosmological observations. These results demonstrate that a minimal and predictive $SO(10)$ scenario with spontaneous $CP$ violation can simultaneously address several outstanding puzzles in particle physics and cosmology.

%%%%%%%%%%%%%%%%%%%%%%%%%%%%%%%%%%%%%%%%%%%%%%%%%
%%%%%%%%%%%%%%%%%%%%%%%%%%%%%%%%%%%%%%%%%%%%%%%%%
\subsection*{Acknowledgments}
The work of KSB is supported in part by the U.S. Department of Energy under grant number DE-SC0016013. 
CSF acknowledges the support by Fundacão de Amparo à Pesquisa do Estado de São Paulo (FAPESP) Contract No. 2019/11197-6  and Conselho Nacional de Desenvolvimento Científico e Tecnológico (CNPq) under Contract No. 304917/2023-0.  SS also acknowledges the financial support
from the Slovenian Research Agency (research core funding No.\ P1-0035 and N1-0321). The authors acknowledge the Center for Theoretical Underground Physics and Related Areas (CETUP* 2024) and the Institute for Underground Science at Sanford Underground Research Facility (SURF) for hospitality and for providing a conducive environment where this work was initiated.  Additionally, KB and CSF acknowledge CETUP* 2025 and the Institute for hospitality when the work was finalized.

\appendix
%%%%%%%%%%%%%%%%%%%%%%%%%%%%%%%%%%%%%%%%%%%%%%%%%%
%%%%%%%%%%%%%%%%%%%%%%%%%%%%%%%%%%%%%%%%%%%%%%%%%%
\section{Some details on the Higgs doublet masses}\label{app:mass_matrix}
In this appendix we provide some details of the relations between bases and Clebsch--Gordan coefficients used to construct the Higgs doublet mass matrix of Eq.~\eqref{eq:doublet}.

%%%%%%%%%%%%%%%%%%%%%%%%%%%%%%%%%%%%%%%%%
\noindent
(i) \underline{First}, we explain how the relations quoted in Eqs.~\eqref{eq:basis01}--\eqref{eq:basis02} that helps to go from one basis to another are derived. For this purpose, we examine the operator $10 \cdot 120 \cdot 45= (10 \cdot 120)_{45} \; \cdot  45$. Although our model does not contain any Higgs field in the $45_H$ representation (which is a two index antisymmetric tensor), it is convenient to use the following trick to derive these relations.
\begin{itemize}
\item  Note that a $45_H$ Higgs can take a general VEV of the form $\langle 45_H \rangle = \mathrm{diag}(\omega_R, \omega_R, \omega_Y, \omega_Y, \omega_Y)\otimes i\tau_2$ along the two SM singlet directions contained in $45_H$. Special cases correspond to
\begin{align}
\omega_R=\omega_Y: &\quad  SO(10)\to SU(5)\times U(1)_X,
\\
\omega_R=0: &\quad  SO(10)\to SU(3)_{c}\times SU(2)_{L}  \times SU(2)_{R}  \times U(1)_{B-L},
\\
\omega_Y=0: &\quad  SO(10)\to SU(4)_{C}\times SU(2)_{L}  \times U(1)_{R}.
\end{align}

\item Noting from Eq.~\eqref{eq:10decompose:S5} that  $10_H$ contains Higgs doublets only in the representations $(5,2)+(\overline 5,-2)$ under the $SU(5)\times U(1)_X$, utilizing the $\omega_R=\omega_Y=\omega$ VEV structure for a $45_H$, the operator under consideration leads to mass term for only $\omega H^{10}_{5} H^{120}_{\overline 5}+c.c$. Knowing the Higgs doublet components within $10_H$, in a certain basis~\cite{Wilczek:1981iz},
\begin{align}
H^{10}_{5}= ( \mathrm{Re}H^+, \mathrm{Im}H^+, \mathrm{Re}H^0, \mathrm{Im}H^0 )^T,    
\end{align}
we identify the Higgs doublet, namely, $H^{120}_{5}$ residing in $120_H\equiv \Phi$,
\begin{align}
H^{120}_{5}= \frac{1}{2} \begin{pmatrix}
\Phi_{341}+\Phi_{561}+\Phi_{781}+\Phi_{901}
\\
\Phi_{342}+\Phi_{562}+\Phi_{782}+\Phi_{902}
\\
\Phi_{123}+\Phi_{563}+\Phi_{783}+\Phi_{903}
\\
\Phi_{124}+\Phi_{564}+\Phi_{784}+\Phi_{904}
\end{pmatrix},    
\end{align}
where the index $0$ is a shorthand for the index $10$.

\item Now, considering the VEV structure of $45_H$ with $\omega_R=0$, which is achieved via the non-zero VEV of the sub-multiplet $(1,1,15)\subset 45_H$, the same operator leads to a mass term of the form $\omega_Y H^{10}_{(2,2,1)} H^{120}_{(2,2,15)}$. Therefore, we identify the components of  $H^{120}_{(2,2,15)}$ as
\begin{align}
H^{120}_{(2,2,15)} = \frac{1}{\sqrt{3}} \begin{pmatrix}
\Phi_{561}+\Phi_{781}+\Phi_{901}
\\
\Phi_{562}+\Phi_{782}+\Phi_{902}
\\
\Phi_{563}+\Phi_{783}+\Phi_{903}
\\
\Phi_{564}+\Phi_{784}+\Phi_{904}
\end{pmatrix}.   
\end{align}

\item Moreover, considering the case with $\omega_Y=0$, which is achieved via the non-zero VEV of the sub-multiplet $(1,3,1)\subset 45_H$, the operator mentioned above leads to a mass term of the form $\omega_R H^{10}_{(2,2,1)} H^{120}_{(2,2,1)}$, which lets us identify the components of  $H^{120}_{(2,2,1)}$,
\begin{align}
H^{120}_{(2,2,1)}=  \begin{pmatrix}
\Phi_{341}
\\
\Phi_{342}
\\
\Phi_{123}
\\
\Phi_{124}
\end{pmatrix}.    
\end{align}

\item The last three items above lets us express $H^{120}_{5}$ in terms of  $H^{120}_{(2,2,1)}$ and $H^{120}_{(2,2,15)}$. On the other hand, the orthogonal combination of which determines $H^{120}_{45}$, as shown in Eqs.~\eqref{eq:basis01}-~\eqref{eq:basis02}.
\end{itemize}

%%%%%%%%%%%%%%%%%%%%%%%%%%%%%%%%%%%%%%%%%
\noindent
(ii) \underline{Second}, in the following, we briefly mention how the relative Clebsch--Gordan factor of \(1/\sqrt{3}\) between \(H^{120}_{\overline{5}} H^{\overline{126}}_{5}\) and \(H^{120}_{45} H^{\overline{126}}_{\overline{45}}\) is obtained.
\begin{itemize}
\item  We now concentrate on the operator of the form $120\cdot 126 \cdot 45= (120\cdot 126)_{45}\; \cdot 45$ and set $\omega_R=\omega_Y=\omega$. As can be seen from Eqs.~\eqref{eq:120decompose:S5}-\eqref{eq:126decompose:S5}, this operator leads to a mixed term, which contains   $\omega H^{120}_{5}H^{126}_{\overline 5}$ and $\omega H^{120}_{\overline{45}}H^{126}_{45}$.

\item  From the item (i) above, we have already identified components of $H^{120}_{5}$ and $H^{120}_{45}$. By utilizing this information, after some tedious algebra, from the operator $120\cdot 126 \cdot 45$, we recognize the fields $H^{126}_{\overline 5}$ and $H^{126}_{45}$. 

\item Upon normalizing these fields properly, we finally obtain this mass mixing term to have the form: $H^{120}_{5}H^{126}_{\overline 5} + (1/\sqrt{3}) H^{120}_{\overline{45}}H^{126}_{45}$.
\end{itemize}

%%%%%%%%%%%%%%%%%%%%%%%%%%%%%%%%%%%%%%%%%
\noindent
(ii) \underline{Third}, we describe the procedure to find the relative Clebsch--Gordan factor in the doublet mass matrix between \(|H^{120}_{(2,2,1)}|^2\) and \(|H^{120}_{(2,2,15)}|^2\). 
\begin{itemize}
\item To determine this Clebsch--Gordan coefficient, we analyze the term 
\((120\,120)_{54} \,(54\,54)_{54}\). Upon inserting the VEV of the field $54_H$, which aligns along the Pati-Salam singlet direction (1,1,1) as shown in Eq.~\eqref{eq:basis02}, this term generates mass contributions of the form 
\(\omega_S^2 |H^{120}_{(2,2,1)}|^2\) and \(\omega_S^2 |H^{120}_{(2,2,15)}|^2\).

\item As the component fields for both $H^{120}_{(2,2,1)}$ and $H^{120}_{(2,2,15)}$ have already been identified in this appendix, substituting them explicitly allows one to compute the relative Clebsch--Gordan factor between these terms, which is found to be \(9/17\).
\end{itemize}

%%%%%%%%%%%%%%%%%%%%%%%%%%%%%%%%%%%%%%%%%%%%%%%%%%
%%%%%%%%%%%%%%%%%%%%%%%%%%%%%%%%%%%%%%%%%%%%%%%%%%
\section{Benchmark fit parameters}\label{app:bestfit}
In this appendix, we present the parameter set corresponding to the benchmark best fit discussed in the main text. These parameters are given as follows.
\begin{align}
&(r_1, r_2) = ( 5.99229\times 10^{-3} e^{i 2.00615}, 1.7078 e^{i 3.0608}),  
\\
&(\alpha, \beta)=(-3.37921, -1.55318)\times 10^{-2},  \quad c_R= 5.38027\times 10^{12}, 
\\
&S= \left(
\begin{array}{ccc}
 1.67119\times 10^{-8} & 0 & 0 \\
 0 & 2.95707\times 10^{-1} & 0 \\
 0 & 0 & 83.8074 \\
\end{array}
\right) \;\rm{GeV},
\\
&D=- \begin{pmatrix}
4.47091 \times 10^{-4} & -2.03111 \times 10^{-3} & 1.77378 \times 10^{-2} \\
-2.03111 \times 10^{-3} & 2.85185 \times 10^{-2} & 3.91335 \times 10^{-1} \\
1.77378 \times 10^{-2} & 3.91335 \times 10^{-1} & 4.80425 \times 10^{-1} \\
\end{pmatrix}
 \;\rm{GeV},
\\
&A=
\begin{pmatrix}
0 & 8.02993 \times 10^{-4} & -1.05388 \times 10^{-2} \\
-8.02993 \times 10^{-4} & 0 & -3.42708 \times 10^{-1} \\
1.05388 \times 10^{-2} & 3.42708 \times 10^{-1} & 0 \\
\end{pmatrix}
\;\rm{GeV}. 
\end{align}

At the renormalization scale \(M_2\), the charged lepton and Dirac neutrino Yukawa couplings—relevant for computing the baryon asymmetry parameter—are expressed in the basis where the right-handed neutrino mass matrix is diagonal as follows:
\begin{align}
&
\left(M_1, M_2, M_3\right)= \left( 8.99144\times 10^4, 1.59098\times 10^{12}, 4.1136\times 10^{14} \right)\; \mathrm{GeV}, \label{eq:117}
\\
&y_{\nu_D}=\scalemath{0.78}
{
\left(
\begin{array}{ccc}
 -2.4594 \times 10^{-6} + 8.3093 \times 10^{-8}i & 3.6491 \times 10^{-6} - 9.8773 \times 10^{-7}i & 1.1392 \times 10^{-6} + 1.1036 \times 10^{-5}i \\
 1.8690 \times 10^{-5} + 2.3009 \times 10^{-7}i & -5.0391 \times 10^{-3} + 5.2654 \times 10^{-6}i & 1.0349 \times 10^{-3} + 3.2502 \times 10^{-4}i \\
 -1.8300 \times 10^{-4} - 4.3011 \times 10^{-6}i & -5.0003 \times 10^{-3} - 1.7226 \times 10^{-4}i & -1.2926 + 8.3277 \times 10^{-5}i \\
\end{array}
\right)
,
}
\\
&y_E=\scalemath{0.78}
{
\left(
\begin{array}{ccc}
 -2.5426 \times 10^{-6} - 8.6071 \times 10^{-8}i & 3.7735 \times 10^{-6} + 1.0170 \times 10^{-6}i & 1.2925 \times 10^{-6} - 1.2507 \times 10^{-5}i \\
 1.93397 \times 10^{-5} - 2.4231 \times 10^{-7}i & -1.4912 \times 10^{-4} - 3.2989 \times 10^{-5}i & 1.1719 \times 10^{-3} - 3.6852 \times 10^{-4}i \\
 -2.0294 \times 10^{-4} + 4.7758 \times 10^{-6}i & -5.5446 \times 10^{-3} + 1.9138 \times 10^{-4}i & 9.4468 \times 10^{-4} - 8.4259 \times 10^{-3}i \\
\end{array}
\right)
}.    
\end{align}
On the other hand, at the renormalization scale \(M_1\), these matrices are as follows: 
\begin{align}
&y_{\nu_D}=\scalemath{0.78}
{
\left(
\begin{array}{ccc}
 -2.3854 \times 10^{-6} + 8.0593 \times 10^{-8}i & 3.5395 \times 10^{-6} - 9.5805 \times 10^{-7}i & 1.1050 \times 10^{-6} + 1.0704 \times 10^{-5}i \\
 1.8690 \times 10^{-5} + 2.3009 \times 10^{-7}i & -5.0391 \times 10^{-3} + 5.2654 \times 10^{-6}i & 1.0349 \times 10^{-3} + 3.2502 \times 10^{-4}i \\
 -1.8300 \times 10^{-4} - 4.3011 \times 10^{-6}i & -5.0003 \times 10^{-3} - 1.7226 \times 10^{-4}i & -1.2926 + 8.3277 \times 10^{-5}i \\
\end{array}
\right)
,
}
\\
&y_E=\scalemath{0.78}
{
\left(
\begin{array}{ccc}
 -2.5896 \times 10^{-6} - 8.7663 \times 10^{-8}i & 3.8434 \times 10^{-6} + 1.0358 \times 10^{-6}i & 1.3164 \times 10^{-6} - 1.2738 \times 10^{-5}i \\
 1.9698 \times 10^{-5} - 2.4647 \times 10^{-7}i & -1.5187 \times 10^{-4} - 3.3590 \times 10^{-5}i & 1.1936 \times 10^{-3} - 3.7533 \times 10^{-4}i \\
 -2.0669 \times 10^{-4} + 4.8641 \times 10^{-6}i & -5.6471 \times 10^{-3} + 1.9492 \times 10^{-4}i & 9.6214 \times 10^{-4} - 8.5816 \times 10^{-3}i \\
\end{array}
\right)
}.
\end{align}
Using these Yukawa coupling, we obtain the baryon asymmetry: $\eta_B = 8.2 (123)  \times 10^{-10}$ for zero (thermal) initial $N_i$ abundance.

\bibliographystyle{style}
\bibliography{reference}
\end{document}